\documentclass{article}
\usepackage{amssymb}
\usepackage{amsmath}
\usepackage{amsthm}
\usepackage{mathtools}
\usepackage{latexsym}
\usepackage{amstext}
\usepackage{xspace}
\usepackage{enumerate}
\usepackage{macrosLLNCS}
\usepackage{graphicx}
\usepackage{fullpage}
\usepackage{color}
\usepackage{stmaryrd}
\usepackage{verbatim}
\usepackage{algorithmic}
\usepackage[ruled,vlined]{algorithm2e}\usepackage{url}

\def \Rm#1{\mbox{\rm #1}}
\def \lsem      {\raise1pt\hbox{\Rm {[\kern-.12em[}}}
\def \rsem      {\raise1pt\hbox{\Rm {]\kern-.12em]}}}

\def \Rm#1{\mbox{\rm #1}}
\def \lsem      {\raise1pt\hbox{\Rm {[\kern-.12em[}}}
\def \rsem      {\raise1pt\hbox{\Rm {]\kern-.12em]}}}

\newcommand {\qc}[1] {{\sf{#1}}}
\def\>{\ensuremath{\rangle}}
\def\<{\ensuremath{\langle}}
\def\sl {\ensuremath{\llparenthesis}}
\def\sr{\ensuremath{\rrparenthesis}}
\def\-{\ensuremath{\textrm{-}}}
\def\ott{t}
\def\otu{u}
\def\ots{s}

\def\ctp{P}
\def\ctq{Q}
\def\fdmu{\Delta}
\def\fdnu{\Xi}
\def\fdomega{\Theta}

\def\dmu{\mu}
\def\dnu{\nu}
\def\domega{\omega}

\def\h{\ensuremath{\mathcal{H}}}
\def\p{\ensuremath{\mathcal{P}}}
\def\l{\ensuremath{\mathcal{L}}}
\def\g{\ensuremath{\mathcal{G}}}
\def\lh{\ensuremath{\mathcal{L(H)}}}
\def\dh{\ensuremath{\mathcal{D(H})}}

\def\r{\ensuremath{\mathcal{R}}}

\def\u{\ensuremath{\mathcal{U}}}
\def\k{\ensuremath{\mathcal{K}}}
\def\K{\ensuremath{\mathfrak{K}}}
\def\S{\ensuremath{\mathfrak{S}}}
\def\s{\ensuremath{\mathcal{S}}}
\def\t{\ensuremath{\mathcal{T}}}
\def\u{\ensuremath{\mathcal{U}}}
\def\U{\ensuremath{\mathfrak{U}}}
\def\L{\ensuremath{\mathfrak{L}}}
\def\x{\ensuremath{\mathcal{X}}}
\def\y{\ensuremath{\mathcal{Y}}}
\def\z{\ensuremath{\mathcal{Z}}}

\def\st{\ensuremath{\mathfrak{t}}}
\def\su{\ensuremath{\mathfrak{u}}}

\def\ra{\ensuremath{\rightarrow}}
\def\a{\ensuremath{\mathcal{A}}}
\def\b{\ensuremath{\mathcal{B}}}

\def\e{\ensuremath{\mathcal{E}}}
\def\f{\ensuremath{\mathcal{F}}}
\def\l{\ensuremath{\mathcal{L}}}

\def\c{\ensuremath{\mathcal{C}}}
\def\d{\ensuremath{\mathcal{D}}}
\def\dh{\ensuremath{\mathcal{D(H)}}}
\def\lh{\ensuremath{\mathcal{L(H)}}}
\def\le{\ensuremath{\sqsubseteq}}
\def\eval{\ensuremath{{\psi}}}

\def\osnt{\ensuremath{\sl \ott, \e\sr}}
\def\snt{\st}

\def\osnu{\ensuremath{\sl \otu, \f\sr}}
\def\osns{\ensuremath{\sl s, \g\sr}}
\def\snu{\su}
\def\fdist{\ensuremath{\d ist_\h}}
\def\dist{\ensuremath{Dist}}

\def\sact{\ensuremath{\gamma}}

\newcommand{\supp}[1]{\ensuremath{\lceil{#1}\rceil}}

\newcommand {\nil} {\mbox{\bf{nil}}}
\newcommand {\iif} {\mbox{\bf{if}}}
\newcommand {\then} {\mbox{\bf{then}}}

\newcommand {\true} {\mbox{\texttt{tt}}}
\newcommand {\false} {\mbox{\texttt{ff}}}

\newcommand{\tr}{{\rm tr}}
\newcommand{\rto}[1]{\stackrel{#1}\longrightarrow}

\newcommand{\srto}[1]{\stackrel{#1}\longmapsto}

\newcommand{\Rto}[1]{\stackrel{#1}\Longrightarrow}

\newcommand{\define}{\stackrel{def}=}

\newcommand{\stbis}{\ \dot\sim\ } 
\newcommand{\nssbis}{\ \dot\nsim\ } 

\newcommand{\bis}{\sim}

\newcommand{\id}{\mathcal{I}}
\newcommand{\stet}[1]{\{ {#1}  \}  } 

\newtheorem{theorem}{Theorem}[section]
\newtheorem{prop}[theorem]{Proposition}
\newtheorem{corollary}[theorem]{Corollary}
\newtheorem{lemma}[theorem]{Lemma}

\newtheorem{definition}[theorem]{Definition}
\newtheorem{example}[theorem]{Example}

\begin{document}%

\title{Symbolic bisimulation for quantum processes}
\author{Yuan Feng$^1$, Yuxin Deng$^2$, and Mingsheng Ying$^1$\\
\\
$^1$ University of Technology, Sydney, Australia, and Tsinghua University, China\\
$^2$Shanghai Jiao Tong University, China
}

\maketitle

\begin{abstract} With the previous notions of bisimulation presented
  in literature, to check if two quantum processes are bisimilar, we
  have to instantiate the free quantum variables of them with
  arbitrary quantum states, and verify the bisimilarity of resultant
  configurations. This makes checking bisimilarity
  infeasible from an algorithmic point of view, because quantum states constitute a continuum. In this paper, we introduce a \textit{symbolic} operational semantics for quantum processes directly at the quantum operation level, which allows us to describe the bisimulation between quantum processes without resorting to quantum states. 
We show that the symbolic bisimulation defined here is equivalent to the open bisimulation for quantum processes in the previous work, when strong bisimulations are considered.  
An algorithm for checking symbolic ground bisimilarity is presented.
We also give a modal logical characterisation for quantum bisimilarity 
based on an extension of Hennessy-Milner logic to quantum processes.

\end{abstract}
\section{Introduction}
An important issue in quantum process algebra is to discover a quantum generalisation of bisimulation preserved by various process constructs, in particular, parallel composition, where one of  the major differences between classical and quantum systems, namely quantum entanglement, is present. 
Jorrand and Lalire \cite{JL04,La06} defined a branching bisimulation for their \emph{Quantum Process Algebra} (QPAlg), which identifies quantum processes whose associated
graphs have the same branching structure. However, their bisimulation 
cannot always distinguish different quantum operations, as quantum states are only compared when they are input or output. Moreover, the derived bisimilarity is not a congruence; it is not preserved by restriction.
Bisimulation defined in \cite{FDJY07} indeed distinguishes different quantum operations but it
works well only for finite processes. Again, it is not preserved by restriction. 
In \cite{YFDJ09}, a congruent 
bisimulation was proposed for a special model where no classical datum is involved. However, as many important quantum communication protocols such as super-dense coding and teleportation cannot be described in that model, the scope of its application is very limited. 

A general notion of bisimulation for the quantum process algebra qCCS developed by the authors was found in~\cite{FDY11}, which enjoys the following nice features: (1) it is applicable to general models where both classical and
quantum data are involved, and recursion is allowed; (2) it is preserved by all the standard process constructs, including parallel composition; and (3) quantum operations are regarded as invisible, so that they can be combined arbitrarily. Independently, a bisimulation congruence in \emph{Communicating Quantum Processes} (CQP), developed by Gay and Nagarajan~\cite{GN05}, was established by Davidson~\cite{Da11}.  Later on, motivated by~\cite{San96},
an open bisimulation for quantum processes was defined in~\cite{DF11} that makes it possible to separate ground bisimulation and the closedness under super-operator applications, thus providing not only a neater and simpler definition, but also a new technique for proving bisimilarity. 

The various bisimulations defined in the literature, however, have a common shortcoming: they all resort to the instantiation of quantum variables by quantum states. As a result, to check whether or not two processes are bisimilar, we have to accompany them with an arbitrarily chosen quantum states, and check if the resultant configurations are bisimilar. Note that all quantum states constitute a continuum. The verification of bisimilarity is actually infeasible from an algorithmic point of view. The aim of the present paper is to tackle this problem by the powerful symbolic technique~\cite{HL95,BC92}. This paper only considers qCCS, but the ideas and techniques developed here apply to other quantum process algebras.   

As a quantum extension of value-passing CCS, qCCS has both (possibly infinite) classical data domain and (doomed-to-be infinite) quantum data domain.
The possibly infinite classical data set can be dealt with by symbolic bisimulation~\cite{HL95} for classical process algebras directly. However, in qCCS, we are also faced with the additional difficulty caused by the infinity of all quantum states. The current paper solves this problem by introducing super-operator valued distributions, which allows us to fold the operational semantics of qCCS into a symbolic version and provides us with a notion, also called symbolic bisimulation for simplicity, where to check the bisimilarity of two quantum processes, only a finite number of process-superoperator pairs need to be considered, without appealing to quantum states.  
To be specific, we propose
\begin{itemize}
\item  a symbolic operational semantics of qCCS in which quantum processes are described directly by the super-operators they can perform. It also incorporates a symbolic treatment for classical data.
\item a notion of symbolic bisimulation, based on the symbolic operational semantics, as well as an efficient algorithm to check its ground version;
\item the coincidence of  symbolic bisimulation with  the open bisimulation defined in~\cite{DF11}, when strong bisimulation is considered. 
\item a modal characterisation of symbolic bisimulation by a quantum
  logic as an extension of Hennessy-Milner logic.
\end{itemize}

The remainder of the paper is organised as follows. In Section~2, we review some
basic notions from linear algebra and quantum mechanics. The syntax and (ordinary) operational semantics of qCCS 
are presented in Section~3. We also review the definition of open bisimulation presented in~\cite{DF11}. 
Section~4 collects some definitions and properties of the semiring of completely positive super-operators. The notion of super-operator valued distributions, which serves as an extension of probabilistic distributions, is also defined.  
Section~5 is the main part of this paper where we present a symbolic operational semantics of qCCS which describes the execution of quantum processes without resorting to concrete quantum states. Based on it, symbolic bisimulation between quantum processes, which also
incorporates a symbolic treatment for classical data, motivated by symbolic bisimulation for classical processes, is presented and shown to be equivalent to the open bisimulation in Section~3.   
Section~6 is devoted to proposing an algorithm to check symbolic ground bisimulation, which is applicable to reasoning about the correctness of existing quantum communication protocols. In section 7 we propose a modal logic which turns out to be both sound and complete with respect to the symbolic bisimulation.
 We outline the main results in Section 8 and point out some directions for further study. In particular, we suggest the potential application of our results in model checking quantum communication protocols.

\section{Preliminaries}
For convenience of the reader, we briefly recall some basic notions
from linear algebra and quantum theory which are needed in this paper. 
For more details, we refer to \cite{NC00}.

\subsection{Basic linear algebra}
A {\it Hilbert space} $\h$ is a complete vector space equipped with an inner
product $$\langle\cdot|\cdot\rangle:\h\times \h\rightarrow \mathbf{C}$$
such that 
\begin{enumerate}
\item
$\langle\psi|\psi\rangle\geq 0$ for any $|\psi\>\in\h$, with
equality if and only if $|\psi\rangle =0$;
\item
$\langle\phi|\psi\rangle=\langle\psi|\phi\rangle^{\ast}$;
\item
$\langle\phi|\sum_i c_i|\psi_i\rangle=
\sum_i c_i\langle\phi|\psi_i\rangle$,
\end{enumerate}
where $\mathbf{C}$ is the set of complex numbers, and for each
$c\in \mathbf{C}$, $c^{\ast}$ stands for the complex
conjugate of $c$. For any vector $|\psi\rangle\in\h$, its
length $|||\psi\rangle||$ is defined to be
$\sqrt{\langle\psi|\psi\rangle}$, and it is said to be {\it normalized} if
$|||\psi\rangle||=1$. Two vectors $|\psi\>$ and $|\phi\>$ are
{\it orthogonal} if $\<\psi|\phi\>=0$. An {\it orthonormal basis} of a Hilbert
space $\h$ is a basis $\{|i\rangle\}$ where each $|i\>$ is
normalized and any pair of them are orthogonal.

Let $\lh$ be the set of linear operators on $\h$.  For any $A\in
\lh$, $A$ is {\it Hermitian} if $A^\dag=A$ where
$A^\dag$ is the adjoint operator of $A$ such that
$\<\psi|A^\dag|\phi\>=\<\phi|A|\psi\>^*$ for any
$|\psi\>,|\phi\>\in\h$. The fundamental {\it spectral theorem} states that
the set of all normalized eigenvectors of a Hermitian operator in
$\lh$ constitutes an orthonormal basis for $\h$. That is, there exists
a so-called spectral decomposition for each Hermitian $A$ such that
$$A=\sum_i\lambda_i |i\>\<i|=\sum_{\lambda_{i}\in spec(A)}\lambda_i E_i$$
where the set $\{|i\>\}$ constitute an orthonormal basis of $\h$, $spec(A)$ denotes the set of
eigenvalues of $A$,
and $E_i$ is the projector to
the corresponding eigenspace of $\lambda_i$.
A linear operator $A\in \lh$ is {\it unitary} if $A^\dag A=A A^\dag=I_\h$ where $I_\h$ is the
identity operator on $\h$. 
The {\it  trace} of $A$ is defined as $\tr(A)=\sum_i \<i|A|i\>$ for some
given orthonormal basis $\{|i\>\}$ of $\h$. It is worth noting that
trace function is actually independent of the orthonormal basis
selected. It is also easy to check that trace function is linear and
$\tr(AB)=\tr(BA)$ for any operators $A,B\in \lh$.

Let $\h_1$ and $\h_2$ be two Hilbert spaces. Their {\it tensor product} $\h_1\otimes \h_2$ is
defined as a vector space consisting of
linear combinations of the vectors
$|\psi_1\psi_2\rangle=|\psi_1\>|\psi_2\rangle =|\psi_1\>\otimes
|\psi_2\>$ with $|\psi_1\rangle\in \h_1$ and $|\psi_2\rangle\in
\h_2$. Here the tensor product of two vectors is defined by a new
vector such that
$$\left(\sum_i \lambda_i |\psi_i\>\right)\otimes
\left(\sum_j\mu_j|\phi_j\>\right)=\sum_{i,j} \lambda_i\mu_j
|\psi_i\>\otimes |\phi_j\>.$$ Then $\h_1\otimes \h_2$ is also a
Hilbert space where the inner product is defined as the following:
for any $|\psi_1\>,|\phi_1\>\in\h_1$ and $|\psi_2\>,|\phi_2\>\in
\h_2$,
$$\<\psi_1\otimes \psi_2|\phi_1\otimes\phi_2\>=\<\psi_1|\phi_1\>_{\h_1}\<
\psi_2|\phi_2\>_{\h_2}$$ where $\<\cdot|\cdot\>_{\h_i}$ is the inner
product of $\h_i$. For any $A_1\in \mathcal{L}(\h_1)$ and $A_2\in
\mathcal{L}(\h_2)$, $A_1\otimes A_2$ is defined as a linear operator
in $\mathcal{L}(\h_1 \otimes \h_2)$ such that for each
$|\psi_1\rangle \in \h_1$ and $|\psi_2\rangle \in \h_2$,
$$(A_1\otimes A_2)|\psi_1\psi_2\rangle = A_1|\psi_1\rangle\otimes
A_2|\psi_2\rangle.$$  The {\it partial trace} of $A\in\mathcal{L}(\h_1
\otimes \h_2)$ with respected to $\h_1$ is defined as
$\tr_{\h_1}(A)=\sum_i \<i|A|i\>$ where $\{|i\>\}$ is an orthonormal
basis of $\h_1$. Similarly, we can define the partial trace of $A$
with respected to $\h_2$. Partial trace functions are also
independent of the orthonormal basis selected.

Traditionally, a linear operator $\e$ on $\lh$ is called a $super$-$operator$ on $\h$.
A super-operator is said to be 
{\it completely positive} if it maps
positive operators in $\mathcal{L}(\h)$ to positive operators in
$\mathcal{L}(\h)$, and for any auxiliary Hilbert space $\h'$, the
trivially extended operator $\mathcal{I}_{\h'}\otimes \e$ also maps
positive operators in $\mathcal{L(H'\otimes H)}$ to positive
operators in $\mathcal{L(H'\otimes H)}$. Here $\mathcal{I}_{\h'}$ is
the identity operator on $\mathcal{L(H')}$. The elegant and powerful
{\it Kraus representation theorem} \cite{Kr83} of completely positive
super-operators states that a super-operator $\e$ is completely positive
if and only if there are some set of operators $\{E_i : i\in I\}$ with appropriate dimension such that
$$
\e(A)=\sum_{i\in I} E_iA E_i^\dag
$$
for any $A\in \lh$. The operators $E_i$ are called Kraus operators
of $\e$. We abuse the notation slightly by denoting $\e=\{E_i : i\in I\}$.
A super-operator $\e$ is said to be
{\it trace-nonincreasing} if $\tr(\e(A))\leq \tr(A)$ for any positive $A\in \lh$, and
{\it trace-preserving} if the equality always holds. Equivalently, a super-operator is trace-nonincreasing completely positive  (resp. 
trace-preserving completely positive) if and only if its Kraus operators $E_i$ satisfy $\sum_i E_i^\dag E_i\leq I$ (resp. $\sum_i E_i^\dag E_i= I$).
In  this paper, we will
use some well-known (unitary) super-operators listed as follows: the quantum control-not
super-operator $\mathcal{CN}=\{C_N\}$ performed on two qubits where 
$$C_N=\left(%
\begin{array}{cccc}
  1 & 0 & 0 & 0 \\
  0 & 1 & 0 & 0 \\
  0 & 0 & 0 & 1 \\
  0 & 0 & 1 & 0
\end{array}%
\right),$$
the 1-qubit Hadamard super-operator $\h=\{H\}$, and Pauli super-operators
$\sigma^0=\{I_2\},\sigma^1=\{X\},\sigma^2=\{Z\}$, and $\sigma^3=\{Y\}$ where
\[
H=\frac{1}{\sqrt{2}}\left(%
\begin{array}{cc}
  1 & 1 \\
  1 & -1 \\
\end{array}%
\right),\ \  I_2=\left(%
\begin{array}{cc}
  1 & 0 \\
  0 & 1 \\
\end{array}%
\right),
\]

\[
X=\left(%
\begin{array}{cc}
  0 & 1 \\
  1 & 0 \\
\end{array}%
\right),\ Z=\left(%
\begin{array}{cc}
  1 & 0 \\
  0 & -1 \\
\end{array}%
\right),\ Y=\left(%
\begin{array}{cc}
  0 & -i \\
  i & 0 \\
\end{array}%
\right).
\]
We also use the notations $\x, \z$, and $\y$ to denote $\sigma^1,\sigma^2$, and $\sigma^3$, respectively.
\subsection{Basic quantum mechanics}

According to von Neumann's formalism of quantum mechanics
\cite{vN55}, an isolated physical system is associated with a
Hilbert space which is called the {\it state space} of the system. A {\it pure state} of a
quantum system is a normalized vector in its state space, and a
{\it mixed state} is represented by a density operator on the state
space. Here a density operator $\rho$ on Hilbert space $\h$ is a
positive linear operator such that $\tr(\rho)= 1$. 
Another
equivalent representation of density operator is probabilistic
ensemble of pure states. In particular, given an ensemble
$\{(p_i,|\psi_i\rangle)\}$ where $p_i \geq 0$, $\sum_{i}p_i=1$,
and $|\psi_i\rangle$ are pure states, then
$\rho=\sum_{i}p_i[|\psi_i\rangle]$ is a density
operator. Here $[|\psi_i\rangle]$ denotes the abbreviation of
$|\psi_i\>\langle\psi_i|$. Conversely, each density operator can be generated by an
ensemble of pure states in this way.  The set of
density operators on $\h$ can be defined as
$$\dh=\{\ \rho\in\lh\ :\  \rho\mbox{ is positive and } \tr(\rho)=
\mbox{1}\}.$$ 

The state space of a composite system (for example, a quantum system
consisting of many qubits) is the tensor product of the state spaces
of its components. For a mixed state $\rho$ on $\h_1 \otimes \h_2$,
partial traces of $\rho$ have explicit physical meanings: the
density operators $\tr_{\h_1}\rho$ and $\tr_{\h_2}\rho$ are exactly
the reduced quantum states of $\rho$ on the second and the first
component system, respectively. Note that in general, the state of a
composite system cannot be decomposed into tensor product of the
reduced states on its component systems. A well-known example is the
 2-qubit state
$$|\Psi\>=\frac{1}{\sqrt{2}}(|00\>+|11\>)
$$
which appears repeatedly in our examples of this paper. This kind of state is called {\it entangled state}.
To see the strangeness of entanglement, suppose a measurement $M=
\lambda_0[|0\>]+\lambda_1[|1\>]$ is applied on the first qubit
of $|\Psi\>$ (see the following for the definition of
quantum measurements). Then after the measurement, the second qubit will
definitely collapse into state $|0\>$ or $|1\>$ depending on whether
the outcome $\lambda_0$ or $\lambda_1$ is observed. In other words,
the measurement on the first qubit changes the state of the second
qubit in some way. This is an outstanding feature of quantum mechanics
which has no counterpart in classical world, and is the key to many
quantum information processing tasks  such as teleportation
\cite{BB93} and super-dense coding \cite{BW92}.

The evolution of a closed quantum system is described by a unitary
operator on its state space: if the states of the system at times
$t_1$ and $t_2$ are $\rho_1$ and $\rho_2$, respectively, then
$\rho_2=U\rho_1U^{\dag}$ for some unitary operator $U$ which
depends only on $t_1$ and $t_2$. In contrast, the general dynamics which can occur in a physical system is
described by a trace-preserving super-operator on its state space. 
Note that the unitary transformation $U(\rho)=U\rho U^\dag$ is
a trace-preserving super-operator. 

A quantum {\it measurement} is described by a
collection $\{M_m\}$ of measurement operators, where the indices
$m$ refer to the measurement outcomes. It is required that the
measurement operators satisfy the completeness equation
$\sum_{m}M_m^{\dag}M_m=I_\h$. If the system is in state $\rho$, then the probability
that measurement result $m$ occurs is given by
$$p(m)=\tr(M_m^{\dag}M_m\rho),$$ and the state of the post-measurement system
is $M_m\rho M_m^{\dag}/p(m).$ 

A particular case of measurement is {\it projective measurement} which is usually represented by a Hermitian operator.  Let  $M$ be a
Hermitian operator and
\begin{equation}\label{eq:specdec}
M=\sum_{m\in spec(M)}mE_m
\end{equation} 
its spectral decomposition. Obviously, the projectors  $\{E_m:m\in
spec(M)\}$ form a quantum measurement. If the state of a quantum
system is $\rho$, then the probability that result $m$ occurs when
measuring $M$ on the system is $p(m)=\tr(E_m\rho),$ and the
post-measurement state of the system is $E_m\rho E_m/p(m).$
Note that for each outcome $m$, the map $$\e_m(\rho) =
E_m\rho E_m$$
is again a super-operator by Kraus Theorem; it is not
trace-preserving in general.

Let $M$ be a projective measurement with Eq.(\ref{eq:specdec}) its spectral decomposition. We call $M$ non-degenerate if for any $m\in spec(M)$, the corresponding projector $E_{m}$ is 1-dimensional; that is, all eigenvalues of $M$ are non-degenerate. Non-degenerate measurement is obviously a very special case of general quantum measurement. However, when an ancilla system lying at a fixed state is provided, non-degenerate measurements together with unitary operators are sufficient to implement general measurements. 

\section{qCCS: Syntax and Semantics}
In this section, we review the syntax and semantics of a quantum extension of value-passing CCS, called qCCS,
introduced in \cite{FDJY07, YFDJ09, FDY11}, and the definition of open bisimulation between qCCS processes presented in~\cite{DF11}.
\subsection{Syntax}

We assume three types of data in qCCS: \qc{Bool} for booleans,  real numbers \qc {Real}
for classical data, and qubits \qc {Qbt} for quantum data. Let $cVar$, ranged over
by $x,y,\dots$, be the set of classical variables, and $qVar$, ranged over by $q,r,\dots$, the set of
 quantum variables. It is assumed that $cVar$ and $qVar$ are both countably infinite.
 We assume a set $Exp$ of classical data expressions over
\texttt{Real}, which includes $cVar$ as a subset and is ranged over by $e,e',\dots$, and a set of boolean-valued expressions $BExp$, ranged over by $b, b',\dots$, with the usual set of boolean operators $\true$, $\false$,
$\neg$, $\wedge$, $\vee$, and $\ra$. In particular, we let $e\bowtie e'$ be a boolean expression for any $e,e'\in Exp$ and $\bowtie \in \{ >, <, \geq, \leq, =\}$.
We further assume that only classical variables can occur free in both data expressions and boolean expressions.
 Let $cChan$
be the set of classical channel names, ranged over by $c,d,\dots$,
and $qChan$ the set of quantum channel names, ranged over by $\qc
c,\qc d,\dots$. Let $Chan=cChan\cup qChan$. A relabeling function
$f$ is a one to one function from $Chan$ to $Chan$ such that
$f(cChan)\subseteq cChan$ and $f(qChan)\subseteq qChan$.

We often abbreviate the indexed set
$\{q_1,\dots,q_n\}$ to $\widetilde{q}$ when $q_1, \dots,q_n$ are
distinct quantum variables and the dimension $n$ is understood. Sometimes we also use $\widetilde{q}$ to denote
the string $q_1\dots q_n$. 
We assume a set of process constant schemes, ranged over by 
$A, B, \dots$. Assigned to each process constant scheme $A$ there are two non-negative 
integers $ar_c(A)$ and $ar_q(A)$. If $\widetilde{x}$ is a tuple of classical variables with
$|\widetilde{x}|=ar_c(A)$, and $\widetilde{q}$ a tuple of distinct quantum variables with
$|\widetilde{q}|=ar_q(A)$, then $A(\widetilde{x},\widetilde{q})$ is
called a process constant. When $ar_c(A)=ar_q(A)=0$, we also 
denote by $A$ the (unique) process constant produced by $A$.

Based on these notations, the syntax of qCCS terms can be given by the Backus-Naur form as
\begin{eqnarray*}
\ott &::=& \nil\ |\ A(\widetilde{e}, \widetilde{q})\ |\ \alpha.\ott\ |\ \ott+\ott\ |\ \ott\| \ott\ |\ \ott\backslash L\ |\ \ott[f]\ |\ \iif \ b\ \then \ \ott\\
\alpha &::= &\tau\ |\ c?x\  |\ c!e\ |\ \qc c?q\ |\ \qc c!q \ |\ \e[\widetilde{q}]\ |\ M[\widetilde{q};x]
\end{eqnarray*}
where $c\in cChan$, $x\in cVar$, $\qc c\in qChan$,
$q\in qVar$, $\widetilde{q}\subseteq qVar$, $e\in Exp$, $\widetilde{e}\subseteq Exp$, $\tau$ is the silent action,
$A(\widetilde{x}, \widetilde{q})$ is a
process constant, $f$ is a relabeling function, $L\subseteq Chan$,
$b\in BExp$, and $\e$ and $M$ are respectively
a trace-preserving super-operator and a non-degenerate projective measurement applying on the Hilbert
space associated with the systems $\widetilde{q}$. In this paper, we 
assume all super-operators are completely positive.

To exclude quantum processes which are not physically implementable, we also require $q\not\in qv(\ott)$ in $\qc c!q.\ott$ and
$qv(\ott)\cap qv(\otu)=\emptyset$ in $\ott \| \otu$, where for a process term $\ott$, $qv(\ott)$ is the set of its free quantum variables inductively defined as follows:
\[\begin{array}{rclrcl}
qv(\nil) & = & \emptyset & qv(\tau.\ott) & = & qv(\ott)\\
qv(c?x.\ott) & = & qv(\ott) & qv(c!e.\ott) & = & qv(\ott) \\
qv(\qc c?q.\ott) & = & qv(\ott)-\{q\} & qv(\qc c!q.\ott) & = & qv(\ott)\cup\{q\} \\
qv(\e[\widetilde{q}].\ott) & = & qv(\ott)\cup \widetilde{q} &
qv(M[\widetilde{q};x].\ott) & = & qv(\ott) \cup \widetilde{q}\\
qv(\ott+\otu) & = & qv(\ott)\cup qv(\otu) \qquad &
qv(\ott \| \otu) & = & qv(\ott)\cup qv(\otu)\\
qv(\ott[f]) & = & qv(\ott) &
qv(\ott\backslash L) & = & qv(\ott)\\
qv(\iif\ b\ \then\ \ott) & = & qv(\ott) &
qv(A(\widetilde{e}, \widetilde{q})) & = & \widetilde{q}.
\end{array}\]
The notion of
free classical variables in quantum processes, denoted by $fv(\cdot)$, can be defined in the
usual way with the only modification that the quantum measurement prefix
$M[\widetilde{q};x]$ has binding power on $x$. A quantum process term $\ott$
is closed if it contains no free classical variables, $i.e.$,
$fv(\ott)=\emptyset$. 
We let $\t$, ranged over by $\ott, \otu, \cdots$, be the set of all qCCS terms, and $\p$, ranged over by $\ctp, \ctq, \cdots$, the set of closed terms.
To complete the definition of qCCS syntax, we assume that for each process constant 
$A(\widetilde{x}, \widetilde{q})$, there is a defining equation 
$$A(\widetilde{x}, \widetilde{q})\define \ott$$
such that $fv(\ott)\subseteq \widetilde{x}$ and $qv(\ctp)\subseteq \widetilde{q}$. Throughout the paper we implicitly assume the convention that process terms are identified up to $\alpha$-conversion.

The process constructs we give here are quite similar to those in
classical CCS, and they also have similar intuitive meanings: $\nil$
stands for a process which does not perform any action; $c?x$ and $
c!e$ are respectively classical input and classical output, while
$\qc c?q$ and $\qc c!q$ are their quantum counterparts. $\e[\widetilde{q}]$
denotes the action of performing the super-operator $\e$ on the
qubits $\widetilde{q}$ while $M[\widetilde{q};x]$ measures the qubits $\widetilde{q}$
according to $M$ and stores the measurement outcome into the
classical variable $x$. $+$ models nondeterministic choice: $\ott+\otu$
behaves like either $\ott$ or $\otu$ depending on the choice of the
environment. $\|$ denotes the usual parallel composition. The
operators $\backslash L$ and $[f]$ model restriction and relabeling,
respectively: $\ott\backslash L$ behaves like $\ott$ as long as any action
through the channels in $L$ is forbidden, and $\ott[f]$ behaves like
$\ott$ where each channel name is replaced by its image under the
relabeling function $f$. Finally, $\iif\ b\ \then\ \ott$ is the
standard conditional choice where $\ott$ can be executed only if $b$ is
\true.

An evaluation $\eval$ is a function from $cVar$ to \qc{Real}; it can be extended in an obvious way to functions from $Exp$ to $\qc{Real}$ and from $BExp$ to $\{\true, \false\}$, and finally, from $\t$ to $\p$. For simplicity, we still use $\eval$ to denote these extensions. Let $\eval\{v/x\}$ be the evaluation which differs from $\eval$ only in that it maps $x$ to $v$.

\subsection{Transitional semantics}

For each quantum variable $q\in  qVar$, we assume a 2-dimensional 
Hilbert space $\h_q$ to be the
state space of the $q$-system. For any $S\subseteq qVar$,  we denote
$$\h_{S}=\bigotimes_{q\in S} \h_q.$$
In particular, $\h = \h_{qVar}$ is the state space of the whole environment consisting of
all the quantum variables. Note
that $\h$ is a countably-infinite dimensional Hilbert space.

Suppose $\ctp$ is a closed quantum process. A pair of the form
$\<\ctp,\rho\>$ is called a configuration, where $\rho\in \dh$ is a density operator
on $\h$. The set of configurations is denoted by $Con$, and ranged over by $\c,\d,\cdots$. 
Let
\begin{eqnarray*}
Act_c&=&\{\tau\}\cup\{c?v,c!v\ |\ c\in cChan, v\in \qc{Real}\}\cup\{\qc c?r,\qc c!r\ |\ \qc c\in qChan, r\in qVar\}.
\end{eqnarray*}
For each $\alpha\in Act_c$, we define the bound quantum variables $qbv(\alpha)$ of $\alpha$ as 
$qbv(\qc c?r) = \{r\}$ and $qbv(\alpha)=\emptyset$ if $\alpha$ is not a quantum input.
The  channel names used in action $\alpha$ is denoted by $cn(\alpha)$; 
that is, $cn(c?v) = cn(c!v) = \{c\}$, $cn(\qc c?r) = cn(\qc c!r) = \{\qc c\}$, and $cn(\tau)=\emptyset$. We also extend the relabelling function
to $Act_c$ in an obvious way.

\begin{center}
\begin{figure}[t] 
\includegraphics[width=\textwidth]{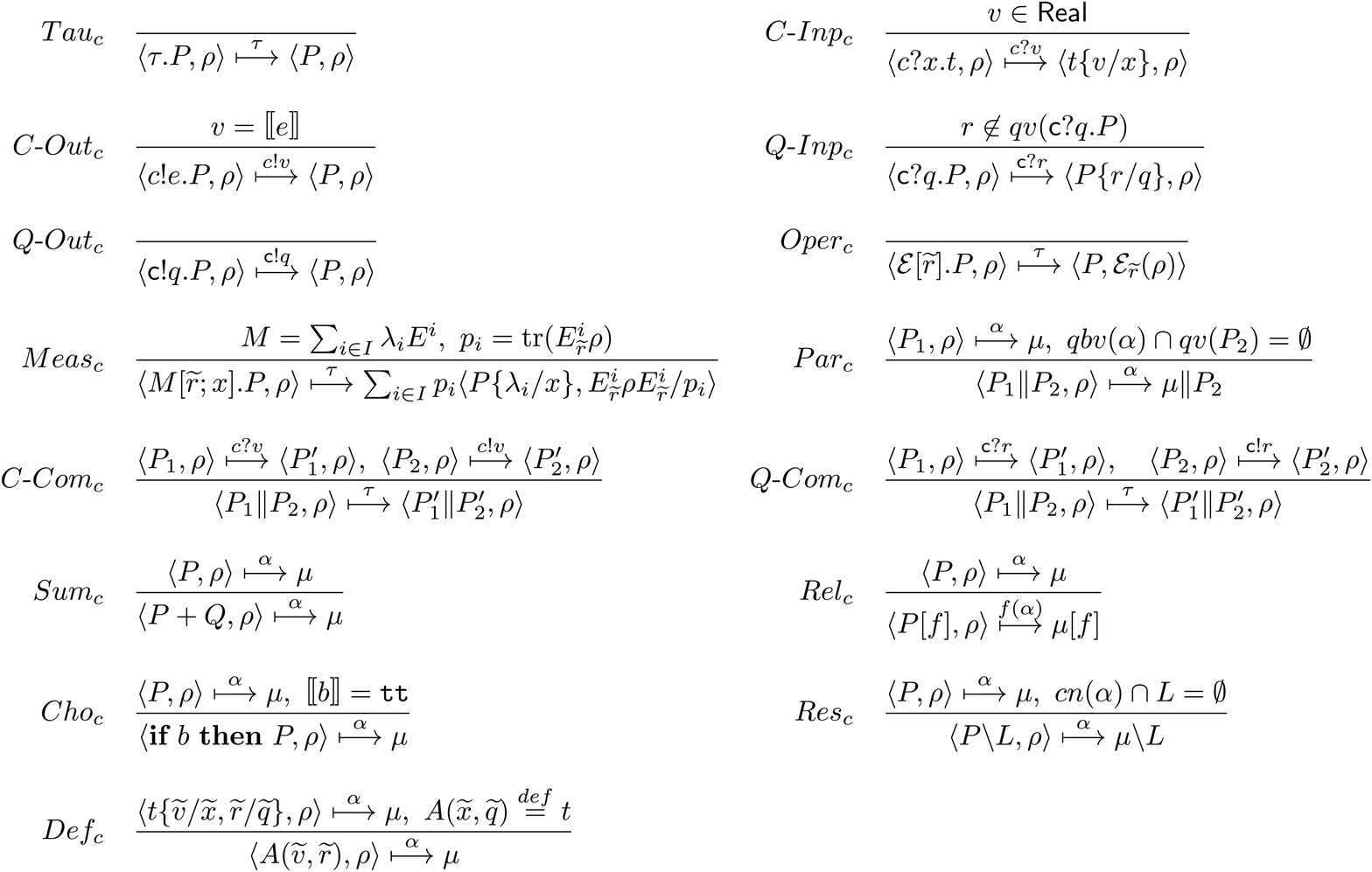}
 \caption{Operational semantics of qCCS \label{fig:csem}}
\end{figure}
\end{center}

Let $\dist(Con)$, ranged over by $\dmu,\dnu,\cdots$, be the set of all finite-supported probabilistic distributions over $Con$. 
Then the operational semantics of qCCS can be
given by the probabilistic labelled transition system (pLTS) 
$\<Con,Act_c,\srto{}\>$, where ${\srto{}}\subseteq Con\times
Act_c\times \dist(Con)$ is the smallest relation satisfying the inference rules
depicted in Fig.~\ref{fig:csem}. 
The symmetric forms for rules $Par_c$, $C\-Com_c$, $Q\-Com_c$, and $Sum_c$ are omitted.

In these rules, we abuse the notation slightly by writing $\c\srto{\alpha}\d$ if $\c\srto{\alpha}\dmu$ where $\mu$ is the simple distribution such that $\dmu(\d)=1$. We also use the obvious extension of the function $\|$ on configurations to distributions. To be precise,
if $\mu=\sum_{i\in I}p_i \<\ctp_i, \rho_i\>$ then $\mu\| \ctq$ denotes the distribution
$\sum_{i\in I}p_i \<\ctp_i\| \ctq, \rho_i\>$.
Similar extension applies to $\dmu[f]$ and $\dmu\backslash L$.


\subsection{Open bisimulation}

In this subsection, we recall the basic definitions and properties of open bisimulation introduced in~\cite{DF11}.
Let $\r\subseteq Con\times Con$ be a relation on configurations. We can lift $\r$ to a relation on $\dist(Con)$ by writing $\dmu\r\dnu$ if 
\begin{enumerate}
\item $\dmu = \sum_{i\in I}p_i
\c_i$, 
\item for each $i\in I$, 
$\c_i\r\d_i$ for some $\d_i$, and 
\item $\dnu=\sum_{i\in I}p_i \d_i$. 
\end{enumerate}
Note that here the set of $\c_i, i\in I,$ are not necessarily distinct.

\begin{definition}\label{def:sbisimulation}
A symmetric relation $\r\subseteq Con\times Con$ is called a
 (strong) open bisimulation if for any $\<\ctp, \rho\>, \<\ctq, \sigma\>\in Con$, $\<\ctp, \rho\>\r \<\ctq, \sigma\>$ implies that 
\begin{enumerate}

\item $qv(\ctp)=qv(\ctq)$, and $\tr_{qv(\ctp)} (\rho) = \tr_{qv(\ctq)} (\sigma)$, 

\item for any trace-preserving super-operator $\e$ acting on $\h_{\overline{qv(\ctp)}}$, whenever $\<\ctp,\e(\rho)\> \srto{\alpha} \dmu$, there
exists $\dnu$ such that $\<\ctq,\e(\sigma)\>\srto{{\alpha}}\dnu$ and
$\dmu\r\dnu$.
\end{enumerate}
\end{definition}

\begin{definition}
\begin{enumerate}
\item Two quantum configurations $\<\ctp, \rho\>$ and $\<\ctq, \sigma\>$ are open bisimilar, denoted by
$\<\ctp, \rho\>\stbis \<\ctq, \sigma\>$, if there exists an open bisimulation $\r$ such that
$\<\ctp, \rho\>\r \<\ctq, \sigma\>$;
\item
Two quantum process terms $\ott$ and $\otu$ are open bisimilar, denoted by
$\ott\stbis \otu$, if for any quantum state $\rho\in \d(\h)$ and any evaluation $\eval$, $\< \ott\eval, \rho\>\stbis \<\otu\eval, \rho \>.$ 

\end{enumerate}
\end{definition}

To illustrate the operational semantics and open bisimulation presented in this section, we give a simple example.
\begin{example}\label{exm:set0c}\rm
This example shows two alternative ways of setting a quantum system to the pure state $|0\>$.
Let $\ctp\define Set^0[q].\id[q].\nil$ and $$\ctq\define M_{0,1}[q;x].(\iif\ x=0\ \then\ \id[q].\nil\ + \iif\ x=1\ \then\ \x[q].\nil),$$
where $Set^0=\{|0\>\<0|, |0\>\<1|\}$, $M_{0,1}$ is the $1$-qubit measurement according to the computational basis $\{|0\>,|1\>\}$, $\id$ is the identity super-operator, and $\x$ is the Pauli-X super-operator.
 For any $\rho\in \dh$, the pLTSs rooted by $\<\ctp, \rho\>$ and $\<\ctq, \rho\>$ respectively are depicted in Fig.~\ref{fig:setzeroc} where 
\begin{eqnarray*}
Q_0&=&\iif\ 0=0\ \then\ \id[q].\nil\ + \iif\ 0=1\ \then\ \x[q].\nil,\\
Q_1&=&\iif\ 1=0\ \then\ \id[q].\nil\ + \iif\ 1=1\ \then\ \x[q].\nil,
\end{eqnarray*}
and $p_i=\tr (|i\>_q\<i|\rho)$.
We can show $P\stbis Q$ by verifying that the relation $\r\cup \r^{-1}$, where
\begin{eqnarray*}
\r=\{(\< P, \rho\>, \<Q, \rho\>), (\< \id[q].\nil, \rho_0\>, \<Q_0, \rho_0\>),  (\< \id[q].\nil, \rho_0\>, \<Q_1, \rho_1\>),  (\< \nil, \rho_0\>, \<\nil, \rho_0\>): \rho\in \dh\}
\end{eqnarray*}
and $\rho_i = |i\>_q\<i|\otimes \tr_q\rho$,
is an open bisimulation.
\begin{figure}[t]
\begin{center}
\includegraphics[width=0.8\textwidth]{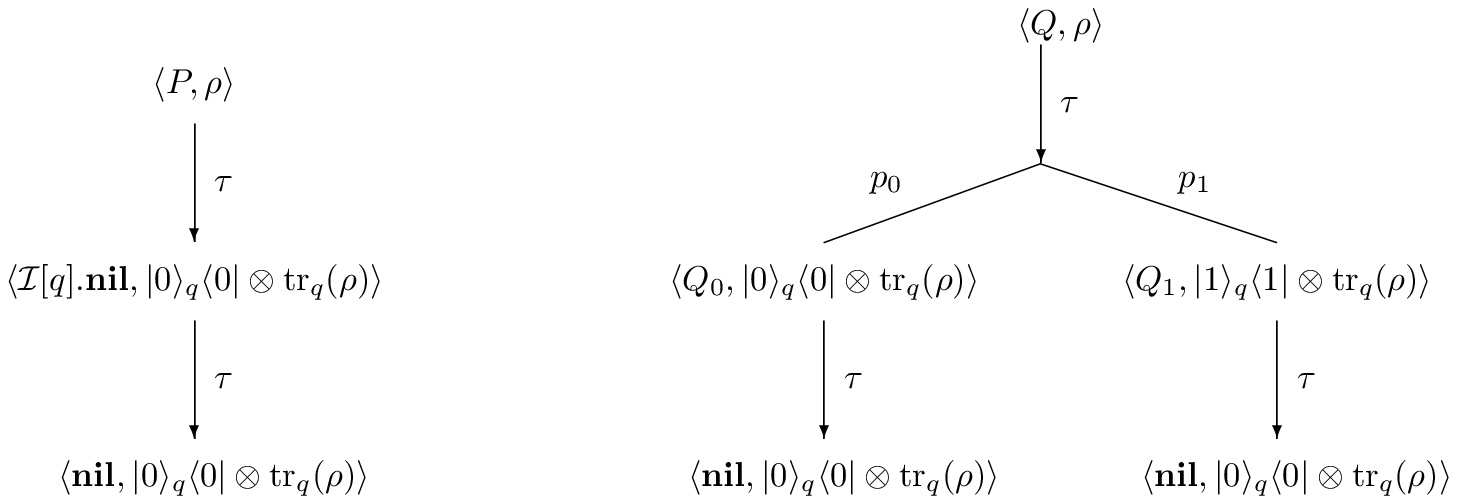}
\end{center}
 \caption{pLTSs for the two ways of setting a quantum system to $|0\>$\label{fig:setzeroc}}
\end{figure}
\end{example}

\section{Super-operator Valued Distributions}

\subsection{Semiring of super-operators} 

We denote by $CP(\h)$ the set of super-operators on $\h$, ranged over by $\a, \b, \cdots$.
Obviously, both $(CP(\h), 0_\h, +)$ and $(CP(\h), \id_\h, \circ)$ are monoids, where $\id_\h$ and $0_\h$ are the identity and null super-operators on $\h$, respectively, and $\circ$ is the composition of super-operators defined by $(\a\circ\b)(\rho) = \a(\b(\rho))$ for any $\rho\in \dh$. We alway omit the symbol $\circ$ and write $\a\b$ directly
for $\a\circ \b$. Furthermore, the operation $\circ$ is (both left and right) distributive with respect to $+$: $$\a(\b_1+\b_2)=\a\b_1 + \a\b_2, \ \ (\b_1+\b_2)\a=\b_1\a + \b_2\a.$$ Thus $(CP(\h), +, \circ)$ forms a semiring.

For any $\a, \b\in CP(\h)$ and $V\subseteq qVar$, we write $\a\lesssim_V \b$ if for any $\rho\in \dh$, $\tr_{\overline{V}}(\a(\rho))\le \tr_{\overline{V}}(\b(\rho))$, where $\overline{V}$ is the 
complement set of $V$ in $qVar$, and $\le$ is the L\"owner preorder defined on operators such as $A\le B$ if and only if $B-A$ is positive semi-definite. Let
$\eqsim_V$ be $\lesssim_V \cap \gtrsim_V$. 
We usually abbreviate $\lesssim_{\emptyset}$ and $\eqsim_{\emptyset}$ to $\lesssim$ and $\eqsim$, respectively. 
It is easy to check that if $\a$ and $\b$ have Kraus operators $\{A_i : i\in I\}$ and $\{B_j : j\in J\}$ respectively, then $\a\lesssim \b$ if and only if $\sum_{i\in I} A_i^\dag A_i \le \sum_{j\in J} B_j^\dag B_j$. The following proposition is direct from definitions:
\begin{prop} Let $\a$ and $\b\in CP(\h)$. Then
\begin{enumerate}
\item $\a\eqsim\id_\h$ if and only if $\a$ is trace-preserving, i.e., $\tr(\a(\rho))=\tr(\rho)$ for any $\rho\in \dh$.
\item $\a\eqsim 0_\h$ if and only if $\a=0_\h$.
\end{enumerate}
\end{prop}

The next lemma, which is easy from definition, shows that the equivalence relation $\eqsim_V$ is preserved by right application of composition.
\begin{lemma}\label{lem:rightapp} Let $\a,\b,\c\in CP(\h)$ and $V\subseteq qVar$. 
If $\a\eqsim_V \b$, then $\a\c\eqsim_V\b\c$.
\end{lemma}

However, $\eqsim$ is not preserved by composition from the left-hand side. A counter-example is when $\a$ is the $X$-pauli super-operator, and $\c$ has one single
Kraus operator $|0\>\<0|$. Then $\a\eqsim \id_\h$, but $\c\a\not\eqsim \c\id_\h$ since $\tr(\c\a(|0\>\<0|))=0$ while  $\tr(\c\id_\h(|0\>\<0|))=1$. Nevertheless, we have the following property which is useful for latter discussion.
\begin{lemma}\label{lem:lrapp} Let $\a,\b\in CP(\h)$ and $\c\in CP(\h_V)$ where $\emptyset\neq V\subseteq qVar$. 
If $\a\eqsim_V \b$, then both $\a\c\eqsim_V\b\c$ and $\c\a\eqsim_V\c\b$.
\end{lemma}
{\it Proof.} Easy from the fact that $\tr_{\overline{V}}\c\a(\rho) = \c(\tr_{\overline{V}}\a(\rho))$ when $\c\in CP(\h_V)$. \hfill $\Box$

Let $CP_t(\h)\subseteq CP(\h)$ be the set of trace-preserving super-operators,
ranged over by $\e, \f, \cdots$. Obviously, $(CP_t(\h), \id_\h, \circ)$ is a sub-monoid of $CP(\h)$ while $(CP_t(\h), 0_\h, +)$ is not. It is easy to check that
for any $\e,\f\in CP_t(\h)$ and $V\subseteq qVar$, $\e\lesssim_V \f$ if and only if $\e\eqsim_V \f$. So for trace-preserving super-operators, we usually use the more symmetric form $\eqsim_V$ instead of $\lesssim_V$.

\subsection{Super-operator valued distributions}

Let $S$ be a countable set.  A super-operator valued distribution, or simply distribution for short, $\fdmu$ over $S$ is a function from $S$ to
$CP(\h)$ such that $\sum_{s\in S}\fdmu(s)\eqsim \id_\h$. We denote by $\supp \fdmu$ the support set of
$\fdmu$, $i.e.$, the set of $s$ such that $\fdmu(s)\neq 0_\h$.
Let $\fdist(S)$ be the set of  finite-support super-operator valued distributions over $S$; that is,
\begin{eqnarray*}
\fdist(S)&=&\{\fdmu:S\rightarrow CP(\h)\ |\
\supp\fdmu \mbox{ is finite, and }\sum_{s\in \supp\fdmu}\fdmu(s)\eqsim \id_\h\}.
\end{eqnarray*}
Let $\fdmu, \fdnu, etc$ range over $\fdist(S)$.
When $\fdmu$ is a simple distribution such that $\supp \fdmu=\{s\}$ for
some $s$ and $\fdmu(s)=\e$, we abuse the notation slightly to denote $\fdmu$ by $\e\bullet s$. We further abbreviate $\id_\h \bullet s$ to $s$.
Note that there are infinitely many different simple distributions having the same support $\{s\}$.

\begin{definition}
Given $\{\fdmu_i : i\in I\}\subseteq \fdist(S)$ and $\{\a_i : i\in I\}\subseteq CP(\h)$, $\sum_{i\in I} \a_i  \eqsim \id_\h$, we define the combination, denoted
by $\sum_{i\in I} \a_i\bullet\fdmu_i$, to be a new distribution $\fdmu$
such that 
\begin{enumerate}
\item $\supp \fdmu = \bigcup\{\supp{\fdmu_i} : i\in I, \a_i\neq 0_\h\}$, 
\item  for any $s\in \supp \fdmu$, $\fdmu(s)=\sum_{i\in I} \fdmu_i(s)\a_i$.
\end{enumerate}
\end{definition}
Here and in the following of this paper, the index sets $I,J,K, etc$ are all assumed to be finite.
By Lemma~\ref{lem:rightapp}, it is easy to check that the above definition is well-defined. Furthermore, since $\eqsim$ is not preserved by left applications of composition, we cannot require $\fdmu(s)=\sum_{i\in I} \a_i\fdmu_i(s)$ in the second clause, although it seems more natural. As a result, say, $\e\bullet (\f\bullet s) = \f\e\bullet s$ but not $\e\f\bullet s$.

Probability distributions can be regarded as special super-operator valued distributions by requiring that all super-operators appeared in the definitions above have the form $p\id_\h$ where $0\leq p\leq 1$. Since in this case all super-operators commute, we always omit the bullet $\bullet$ in the expressions.

\section{Symbolic bisimulation}\label{ssec:opsem}

\subsection{Super-operator weighted transition systems}
\begin{center}
\begin{figure}[t] 
\includegraphics[width=\textwidth]{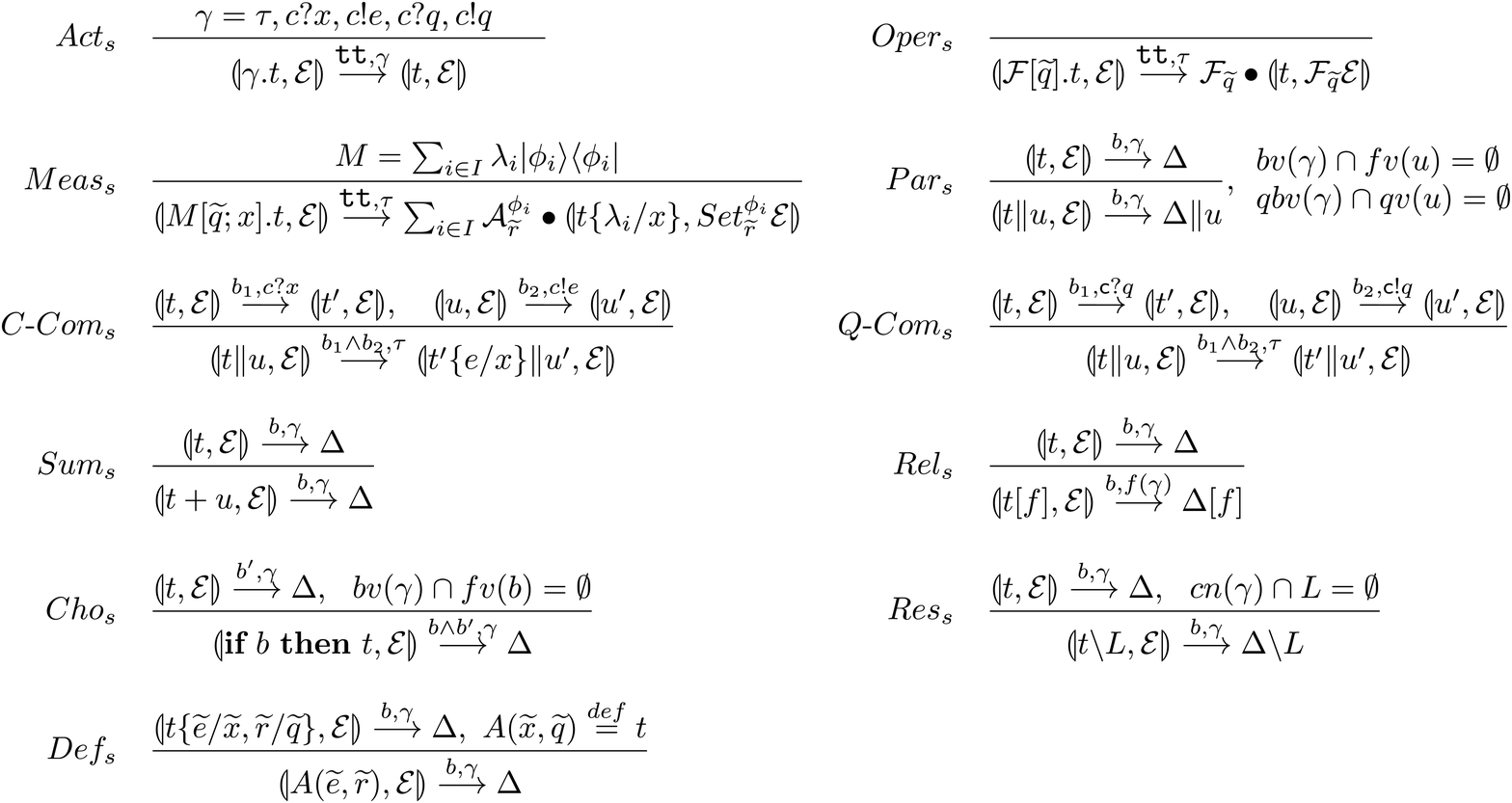}
 \caption{Symbolic operational semantics of qCCS\label{fig:osem}}
\end{figure}
\end{center}

We now extend the ordinary 
probabilistic labelled transition systems to super-operator weighted
ones.
\begin{definition}
A super-operator weighted labelled transition system, or quantum labelled transition system (qLTS), is a triple $(S, Act, \rto{})$, where
\begin{enumerate}
\item $S$ is a countable set of states,
\item $Act$ is a countable set of transition actions,
\item $\rto{}$, called transition relation, is a subset of $S\times Act\times \fdist(S)$.
\end{enumerate}
\end{definition}
For simplicity, we write $s
\rto{\alpha} \fdmu$ instead of $(s, \alpha,\fdmu)\in\rto{}$. A pLTS may be viewed as a degenerate qLTS in which
all super-operator valued distributions are probabilistic ones. 

\subsection{Symbolic transitional semantics of qCCS}\label{subsec:sts}

To present the symbolic operational semantics of quantum processes, we need some more notations. Let
\begin{eqnarray*}
Act_s&=&\{\tau\}\cup\{c?x,c!e\ |\ c\in cChan, x\in cVar, e\in Exp\}\cup\{\qc c?r,\qc c!r\ |\ \qc c\in qChan, r\in qVar\}
\end{eqnarray*}
and $BAct_s=BExp\times Act_s$. For each $\sact\in Act_s$, the notion $qbv(\sact)$ for bound quantum variables, $cn(\sact)$ for channel names, and $fv(\sact)$ for free classical variables are similarly defined as for $Act_c$.
We also define $bv(\sact)$, the set of bound classical variables in $\sact$ in an obvious way.

A pair of the form
$\sl  \ott, \e\sr$, where $t\in \t$ and $\e\in CP_t(\h)$, is called a snapshot, and
 the set of snapshots is denoted by  $SN$. 
Then the symbolic semantics of qCCS is
given by the qLTS
$(SN,BAct_s,\rto{})$ on snapshots, where ${\rto{}}\subseteq SN\times
BAct_s\times \fdist(SN)$ is the smallest relation satisfying the rules
defined in Fig.~\ref{fig:osem}. In Rule $Meas_s$, for each $i\in I$, $\a^{\phi_i}_{\widetilde{r}}\in CP(\h)$ and $Set^{\phi_i}_{\widetilde{r}}\in CP_t(\h)$ are defined respectively as
\begin{eqnarray}
\a^{\phi_i}_{\widetilde{r}}&:& \rho \mapsto  |\phi_i\>_{\widetilde{r}}\< \phi_i|\rho  |\phi_i\>_{\widetilde{r}}\< \phi_i|\label{eq:aphi}\\
Set^{\phi_i}_{\widetilde{r}}&:& \rho \mapsto  \sum_{j\in I} |\phi_i\>_{\widetilde{r}}\< \phi_j|\rho  |\phi_j\>_{\widetilde{r}}\< \phi_i|.\label{eq:seti}
\end{eqnarray}
The symmetric forms for rules $Par_s$, $C\-Com_s$, $Q\-Com_s$, and $Sum_s$ are omitted.
Here again, the functions $\|$, $[f]$, and $\backslash L$ have been extended to super-operator valued distributions by denoting, say,
$\fdmu\| \otu$ the super-operator valued distribution
$\sum_{i\in I}\a_i\bullet \sl\ott_i\|\otu, \e_i\sr$, if $\fdmu=\sum_{i\in I}\a_i\bullet \sl\ott_i, \e_i\sr$.

The transition graph of a snapshot is depicted as usual where
each transition $\osnt\rto{b, \sact}\sum_{i=1}^n \a_i\bullet \sl \ott_i, \e_i\sr$  is
depicted as
\begin{figure}[h] \centering \includegraphics[width=0.5\textwidth]{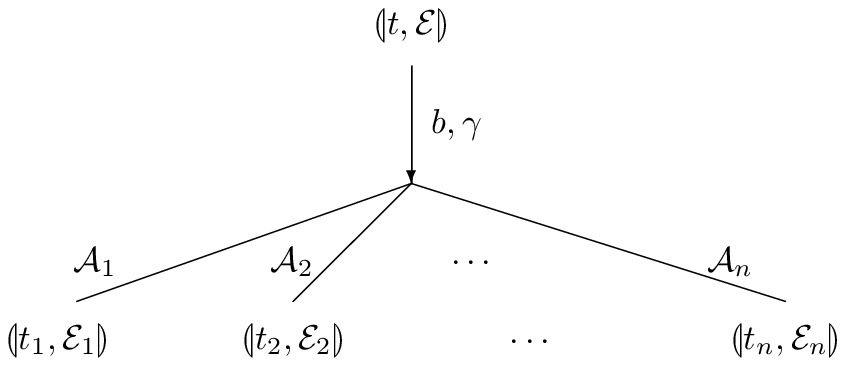}
\end{figure}

\noindent We sometimes omit the line marked with $\id_\h$ for simplicity.

\begin{figure}[t]
\begin{center}
\includegraphics[width=0.8\textwidth]{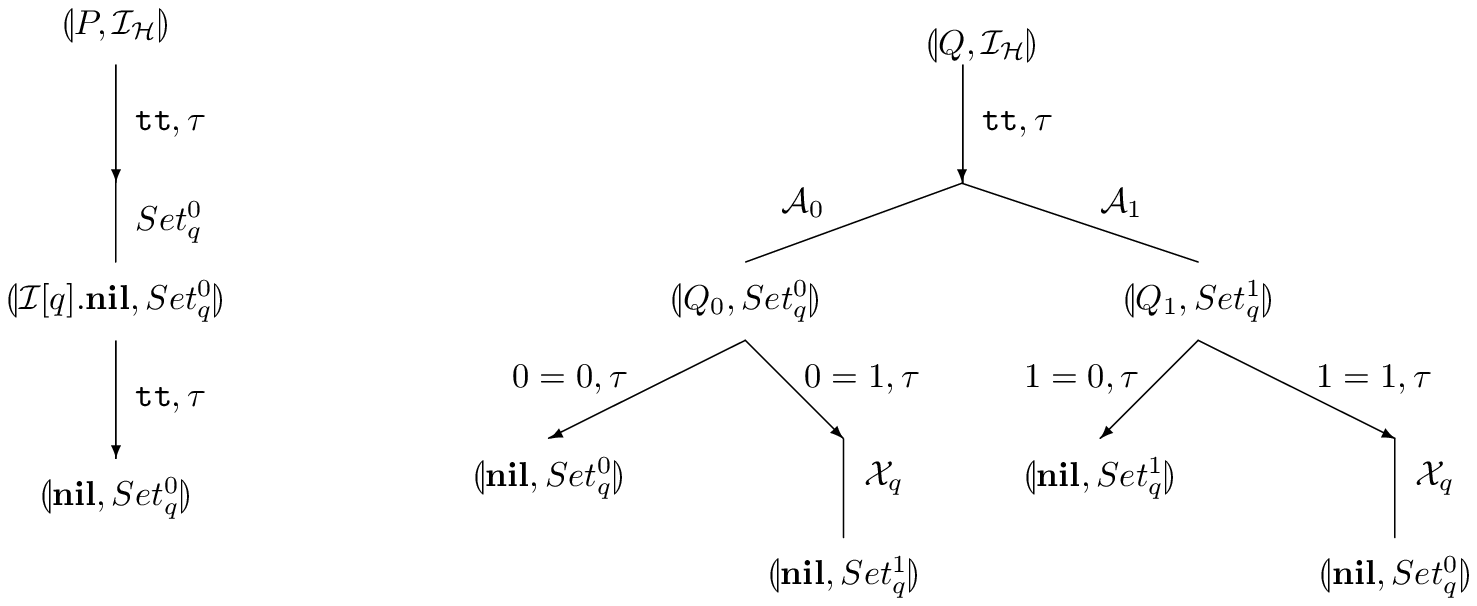}
\end{center}
 \caption{qLTSs for two ways of setting a quantum system to $|0\>$\label{fig:setzero}}
\end{figure}

\begin{example}\label{exm:set0}\rm
(Example~\ref{exm:set0c} revisited) For the first example, we revisit the two ways of setting a quantum system to pure state $|0\>$, presented in Example~\ref{exm:set0c}.
According to the symbolic operational semantics presented in Fig.~\ref{fig:osem}, the qLTSs rooted by $\sl\ctp, \id_\h\sr$ and $\sl\ctq, \id_\h\sr$ respectively can be depicted as in Fig.~\ref{fig:setzero}, where $\a_i$  has the single Kraus operator $|i\>_q\<i|$ for $i=0,1$.

At the first glance, it is tempting to think that symbolic semantics provides no advantage in describing quantum processes, as the qLTSs in Fig.~\ref{fig:setzero} are almost the same as the pLTSs in Fig.~\ref{fig:setzeroc} (Indeed, the right-hand side qLTS in the former is even more complicated than the corresponding pLTS in the latter). However, pLTSs in Fig.~\ref{fig:setzeroc} are depicted for a fixed quantum state $\rho$; to characterise the behaviours of a quantum process, infinitely many such pLTSs must be given, although typically they share the same structure. On the other hand, the qLTSs in Fig.~\ref{fig:setzero} specify $all$ possible behaviours of the processes, by means of the super-operators they can perform.
\end{example}

\begin{figure}[t]
\[
\begin{array}{lr}
\begin{array}{l}
\includegraphics[width=0.45\textwidth]{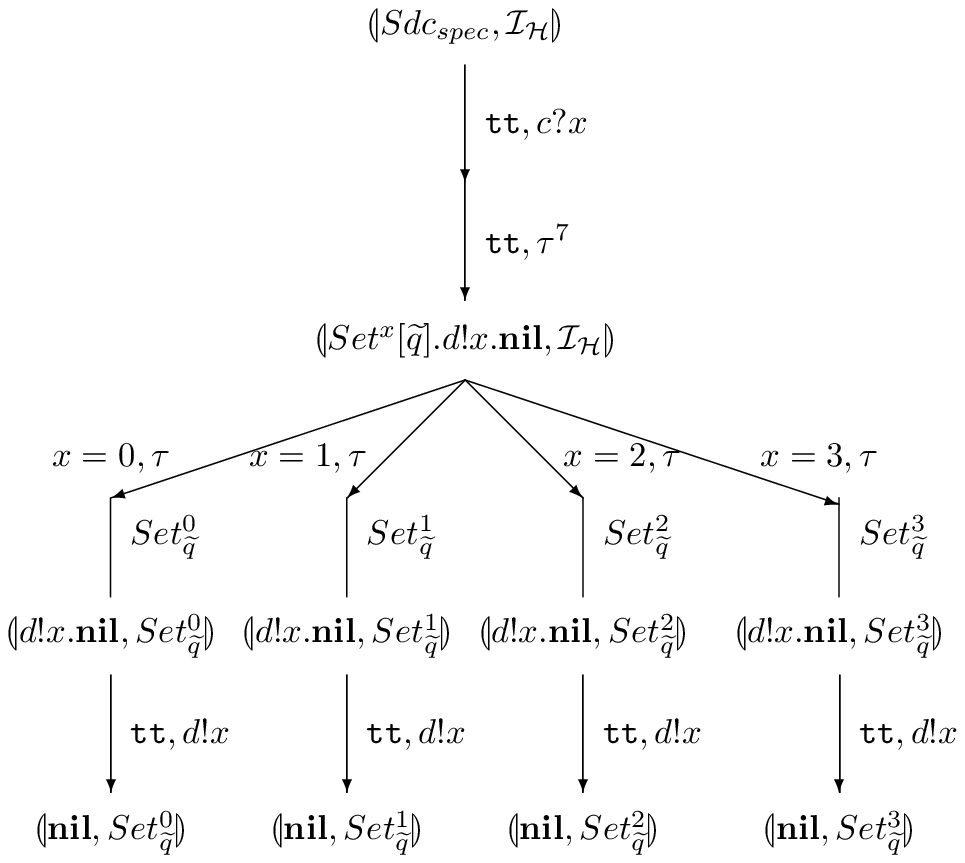}
\end{array}
&
\begin{array}{l}
\includegraphics[width=0.45\textwidth]{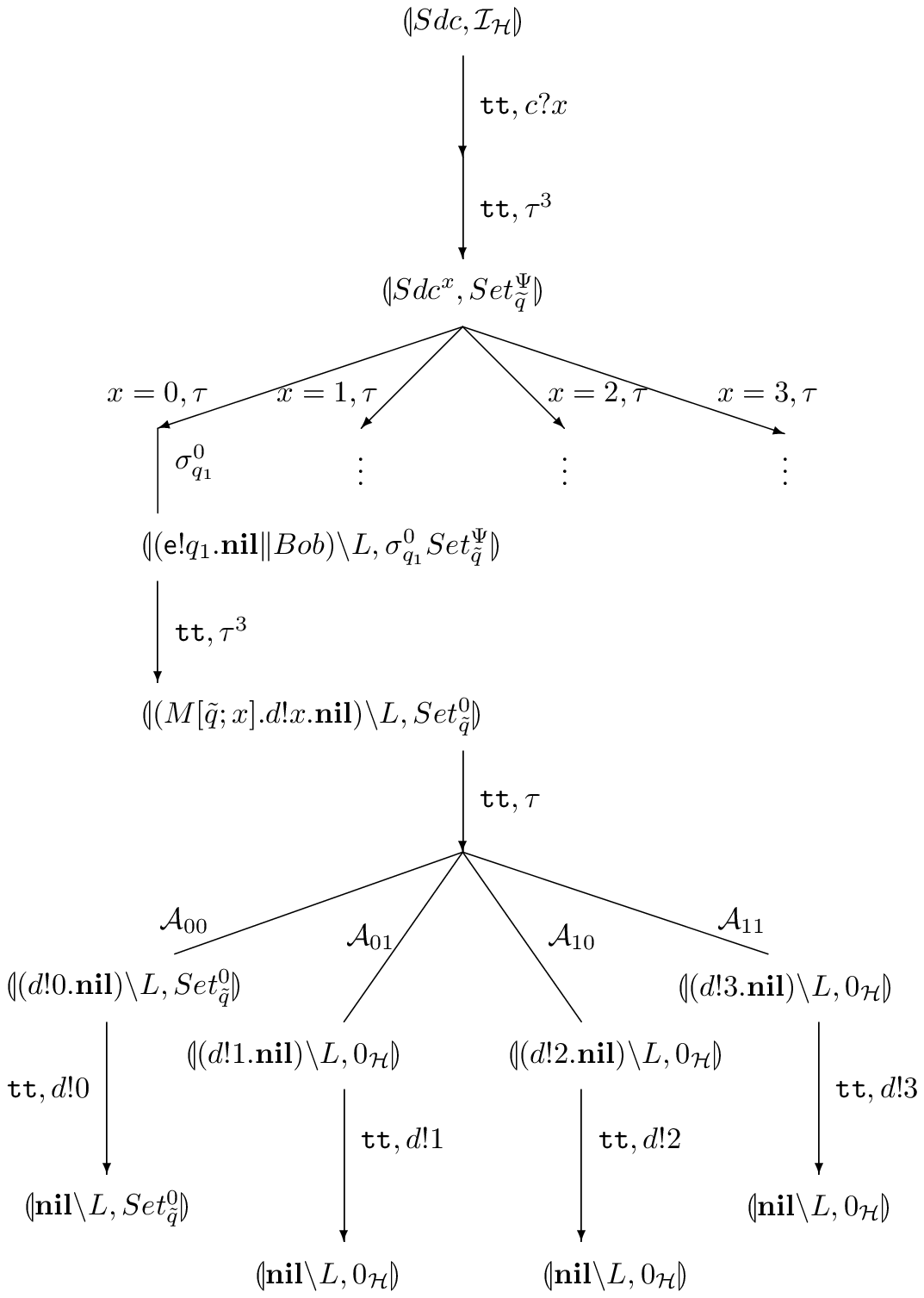}
\end{array}
\end{array}
 \] 
 \caption{qLTSs for $\sl Sdc_{spec}, \id_\h\sr$ and $\sl Sdc, \id_\h\sr$\label{fig:sdc}}
\end{figure}

\begin{example}\label{exm:sdc}\rm This example shows the correctness of super-dense coding protocol.
Let $M=\sum_{i=0}^3
i|\tilde{i}\>\<\tilde{i}|$ be a 2-qubit measurement where $\tilde{i}$ is the binary
expansion of $i$. Let $\mathcal{CN}$ be the controlled-not operation and $\mathcal{H}$ Hadamard operation. Then the quantum processes participating in
super-dense coding protocol can be defined as follows:
\begin{eqnarray*}
Alice &\define& \qc c_A?q_1.\sum_{0\leq i \leq 3}\left(\iif\ x=i\ \then\ \sigma^i[q_1].\qc e!q_1.\nil\right),\\
Bob &\define& \qc c_B?q_2.\qc e?q_1.\mathcal{CN}[q_1,q_2].\h[q_1].M[q_1,q_2; x]. d!x.\nil,\\
EPR &\define& Set^{\Psi}[q_1,q_2].\qc c_B!q_2.\qc c_A!q_1.\nil,\\
Sdc &\define& c?x.(EPR\| Alice\| Bob)\backslash \{\qc c_A, \qc c_B, \qc e\}.
\end{eqnarray*}
The specification of  super-dense coding protocol can be defined as:
$$Sdc_{spec} \define c?x.\tau^7.Set^x[q_1,q_2].d!x.\nil$$
where  $$Set^x[q_1,q_2].d!x.\nil=\sum_{i=0}^3 (\iif\ x=i\ \then\ Set^i[q_1,q_2].d!x.\nil).$$
Here $Set^{i}$ and $Set^{\Psi}$ are the 2-qubit super-operators which set the target qubits to $|\widetilde{i}\>$ and $|\Psi\> = (|00\> + |11\>)/\sqrt{2}$, respectively. 
We insert seven $\tau$'s in the specification to match the internal actions of $Sdc$. The qLTSs rooted from $\sl Sdc_{spec}, \id_\h\sr$ and $\sl Sdc, \id_\h\sr$ respectively are depicted in Fig.~\ref{fig:sdc} where
$\widetilde{q}= \{q_1,q_2\}$, $\a_{\widetilde{i}}$ is the super-operator with the single Kraus operator $|\widetilde{i}\>\<\widetilde{i}|$, $L=\{\qc c_A, \qc c_B, \qc e\}$,
\begin{eqnarray*}
Sdc^x&=&\left(\left(\sum_{i=0}^3(\iif\ x=i\ \then\ \sigma^i[q_1].\qc e!q_1.\nil)\right)\| Bob\right)\backslash\{\qc e\},
\end{eqnarray*}
and for simplicity, we only draw the transitions along the $x=0$ branch.
\end{example}

To conclude this subsection, we prove some useful properties of symbolic transitions.

\begin{lemma}\label{lem:superoperatorc} If
$ \osnt\rto{b, \sact}\fdmu$, then there exist super-operators $\{\b_i : i\in I\}\subseteq CP(\h)$ and $\{\f_i : i\in I\}\subseteq CP_t(\h)$,  and
process terms $\{\ott_i : i\in I\}\subseteq \t$ such that
\begin{enumerate}
\item $\sum_{i\in I}\b_i \eqsim \id_\h$, 
\item $\fdmu=\sum_{i\in I}\b_i \bullet \sl \ott_i, \f_i \e\sr,$
\item for any $\g\in CP_t(\h)$,
$\sl \ott,\g\sr\rto{b,\sact}\sum_{i\in I}\b_i \bullet \sl \ott_i, \f_i \g\sr.$
\end{enumerate}
Especially, if $|I|>1$ then $\b_i$ and $\f_i$ take the forms as $\a^{\phi_i}_{\widetilde{r}}$ and $Set^{\phi_i}_{\widetilde{r}}$ in Eqs.(\ref{eq:aphi}) and (\ref{eq:seti}), respectively.
\end{lemma}
{\it Proof.} Easy from the definition of inference rules. \hfill $\Box$

The following lemmas show the relationship between transitions in ordinary semantics and in symbolic semantics. Let $\eval$ be an evaluation,
$\alpha\in Act_c$, and $\sact\in Act_s$.
We write $\alpha =_\eval \sact$ if either
$\alpha = c!v$, $\sact=c!e$, and $\eval(e)=v$, or $\sact=\alpha$ if neither of them is a classical output.

\begin{lemma}\label{lem:so2s} Suppose $\< \ott\eval, \rho\>\srto{\alpha}\mu$.
Then there exist $b, I$, $\eval'$, 
$\{\a_i : i\in I\}\subseteq CP(\h)$, $\{\e_i : i\in I\}\subseteq CP_t(\h)$,  and
$\{\ott_i : i\in I\}\subseteq \t$,
 such that $\sum_{i\in I}\a_i \eqsim \id_\h$, and
 \begin{enumerate}
 \item $\eval(b)=\true$, 
 \item $\mu=\sum_{i\in I}\tr(\a_{i}(\rho))\< \ott_i\eval',\e_i(\rho)\>$, 
\item for any $\e\in CP_t(\h)$, $\osnt\rto{b, \sact} \sum_{i\in I} \a_i\bullet \sl \ott_i, \e_i\e\sr$, where 
\begin{enumerate}
\item if $\alpha=c?v$ then $\sact=c?x$ for some $x\not\in fv(t)$, and $\eval'=\eval\{v/x\}$,
\item
otherwise, $\sact=_\eval\alpha$ and $\eval'=\eval$.
\end{enumerate}
\end{enumerate}
\end{lemma}
{\it Proof.} We prove by induction on the depth of the inference by which the action $\< \ott\eval, \rho\>\srto{\alpha}\dmu$ is inferred. We argue by cases on the form of $\ott$. 
\begin{enumerate}
\item $\ott=c?x.\ott'$. Then $\ott\eval = c?x.\otu$ where $\otu$ is the process term obtained from $\ott'$ by instantiating all the free variables in $fv(\ott')-\{x\}$ according to $\eval$. By Rule $C\-Inp_c$ we deduce that $\alpha=c?v$ for some $v\in \qc{Real}$ and $\dmu=\<\ctp, \rho\>$ where $\ctp=\otu\{v/x\}= t'\eval\{v/x\}$. By Rule $Act_s$, for any $\e\in CP_t(\h)$, we have $\osnt\rto{\true, c?x}\sl t', \e\sr$. 
So we need only to take $b=\true$, $|I|=1$, $\ott_i=\ott'$, $\a_i=\e_i=\id_\h$.

\item $\ott=c!e.\ott'$. Then $\ott\eval = c!\eval(e).(\ott'\eval)$, and by Rule $C\-Out_c$ we deduce that $\alpha=c!\eval(e)$ and $\dmu=\<\ott'\eval, \rho\>$. 
By Rule $Act_s$, for any $\e\in CP_t(\h)$, we have $\osnt\rto{\true, c!e}\sl \ott', \e\sr$.
 So we need only to take $b=\true$, $|I|=1$, $\ott_i=\ott'$, $\a_i=\e_i=\id_\h$ as well.

\item $\ott=\qc c?q.\ott'$. Then $\ott\eval = \qc c?q.(\ott'\eval)$, and by Rule $Q\-Inp_c$ we deduce that $\alpha=\qc c?r$ for some $r\not\in qv(\ott)$ and $\dmu=\<(\ott'\eval)\{r/q\}, \rho\>$.  By Rule $Act_s$ and $\alpha$-conversion, for any $\e\in CP_t(\h)$, we have $\osnt\rto{\true, \qc c?r}\sl \ott'\{r/q\}, \e\sr$.
 So we need only to take $b=\true$, $|I|=1$, $\ott_i=\ott'\{r/q\}$, $\a_i=\e_i=\id_\h$.

\item $\ott=M[\widetilde{q};x].\ott'$. Then $\ott\eval = M[\widetilde{q};x].\otu$ where $\otu$ is the process term obtained from $\ott'$ by instantiating all the free variables in $fv(\ott')-\{x\}$ according to $\eval$. 
Let $M=\sum_{i\in I}\lambda_i |\phi_i\>\<\phi_i|$.
By Rule $Meas_c$ we deduce that $\alpha=\tau$ and $\dmu=\sum_{i\in I}\tr(\a_{i}(\rho))\< \ctp_i,\e_i(\rho)\>$ where $\ctp_i=\otu\{\lambda_i/x\}= \ott'\{\lambda_i/x\}\eval$, $\a_i=\{|\phi_i\>\<\phi_i|\}$, and $\e_i=\{|\phi_i\>\<\phi_j| : j\in I\}$. Take $b=\true$.
By Rule $Meas_s$, for any $\e\in CP_t(\h)$, we have $\osnt\rto{b, \tau} \sum_{i\in I} \a_i\bullet \sl \ott'\{\lambda_i/x\}, \e_i\e\sr$.

\item $\ott=\ott_1\|\ott_2$. Then $\ott\eval = \ott_1\eval\|\ott_2\eval$. There are two sub-cases to consider:
\begin{enumerate}
\item The action is caused by one of the components, say $\<\ott_1\eval, \rho\>\srto{\alpha}\dmu_1$. Then we have $qbv(\alpha)\cap qv(\ott_2\eval)=\emptyset$, and $\dmu=\dmu_1\|\ott_2\eval$. By induction, there exist $b, I$, $\ott_i$, $\a_i, \e_i$, $i\in I$, such that $\eval(b)=\true$, 
$\mu_1=\sum_{i\in I}\tr(\a_{i}(\rho))\< \ott_i\eval',\e_i(\rho)\>$, and
 for any $\e\in CP_t(\h)$, $\sl \ott_1, \e\sr\rto{b, \sact} \sum_{i\in I} \a_i\bullet \sl \ott_i, \e_i\e\sr$. 
 Note that by $\alpha$-conversion, when $\sact=c?x$, we can always take $x$ such that $x\not\in fv(\ott_2)$, and consequently, $(\ott_i\|\ott_2)\eval'=\ott_i\eval'\|\ott_2\eval$. Finally, we have $\osnt\rto{b, \sact} \sum_{i\in I} \a_i\bullet \sl \ott_i\|\ott_2, \e_i\e\sr$, using Rule $Par_s$. 
 
 \item The action is caused by a (classical or quantum) communication. Here we only detail the case when $\<\ott_1\eval, \rho\>\srto{c?v}\<\ctp_1, \rho\>$, $\<\ott_2\eval, \rho\>\srto{c!v}\<\ctp_2, \rho\>$, $\alpha=\tau$, and $\dmu=\<\ctp_1\|\ctp_2, \rho\>$. Then by induction, there exist $b_1$, $b_2$, $\ott'_1$, $\ott'_2$ such that $\eval(b_1\wedge b_2)=\true$, $\ctp_1=\ott_1'\eval'$, $\ctp_2=\ott_2'\eval$, and for any $\e\in CP_t(\h)$, $\sl \ott_1, \e\sr\rto{b_1, c?x} \sl \ott_1', \e\sr$ and $\sl \ott_2, \e\sr\rto{b_2, c!e} \sl \ott_2', \e\sr$,  where $x\not\in fv(\ott_1)$, $\eval'=\eval\{v/x\}$, and $\eval(e)=v$. Thus
  \begin{eqnarray*}
(\ott_1'\{e/x\}\|\ott_2')\eval=\ott_1'\{e/x\}\eval\|\ott_2'\eval=\ott_1'\eval\{v/x\}\|\ott_2'\eval=\ott_1'\eval'\|\ott_2'\eval=P_1\|P_2.
\end{eqnarray*}
Finally,  we have $\osnt\rto{b_1\wedge b_2, \tau} \sl \ott_1'\{e/x\}\|\ott_2', \e\sr$,  using Rule $Q\-Com_s$. 
\end{enumerate}
\item Other cases. Similar to the cases we discussed above. \hfill $\Box$
\end{enumerate}

\begin{lemma}\label{lem:ss2o} Suppose $\sl \ott, \e\sr\rto{b, \sact} \fdmu$. Then there exist $I$, 
$\{\a_i : i\in I\}\subseteq CP(\h)$, $\{\e_i : i\in I\}\subseteq CP_t(\h)$,  and
$\{\ott_i : i\in I\}\subseteq \t$,
 such that $\sum_{i\in I}\a_i \eqsim \id_\h$, and
 \begin{enumerate}
\item $\fdmu=\sum_{i\in I} \a_i\bullet \sl \ott_i, \e_i\e\sr$, 
\item for any $\eval$ and $\rho$,  $\eval(b)=\true$ implies
$\< \ott\eval, \rho\>\srto{\alpha}\sum_{i\in I}\tr(\a_{i}(\rho))\< \ott_i\eval',\e_i(\rho)\>$
 where  
 \begin{enumerate}
\item
if $\sact=c?x$ then $\alpha=c?v$ for some $v\in \qc{Real}$, and $\eval'=\eval\{v/x\}$,
\item
otherwise, $\sact=_\eval\alpha$ and $\eval'=\eval$.
\end{enumerate}
\end{enumerate}
\end{lemma}
{\it Proof.} Similar to Lemma~\ref{lem:so2s}.\hfill $\Box$

\subsection{Symbolic bisimulation}

Let $\s\subseteq SN\times SN$ be an equivalence relation. We lift $\s$ to $\fdist(SN)\times \fdist(SN)$ by defining $\fdmu\s\fdnu$ if for any equivalence class $T\in SN/\s$, $\fdmu(T)\eqsim\fdnu(T)$; that is, $\sum_{\osnt\in T}\fdmu(\osnt)\eqsim \sum_{\osnt\in T}\fdnu(\osnt)$. We write $\sact =_b \sact'$ if either
$\sact = c!e$, $\sact'=c!e'$, and $b\ra e=e'$, or $\sact=\sact'$ if neither of them is a classical output.

\begin{definition}\label{def:ssbisimulation}
Let $\S=\{\s^b : b\in BExp\}$ be a family of equivalence relations on $SN$. $\S$ is called a symbolic (open) bisimulation if for any $b\in BExp$, $\osnt\s^b\osnu$ implies that 
\begin{enumerate}
\item
$qv(\ott)=qv(\otu)$ and $\e\eqsim_{\overline{qv(\ott)}} \f$, if $b$ is satisfiable;
\item for any $\g\in CP_t(\h_{\overline{qv(\ott)}})$, whenever $\sl\ott,\g\e\sr \rto{b_1, \sact} \fdmu$ with $bv(\sact)\cap fv(b, \ott, \otu)=\emptyset$, then there exists a collection of booleans $B$ such that $b\wedge b_1\ra \bigvee B$ and $\forall\ b'\in B$, $\exists b_2, \sact'$ with $b'\ra b_2$,
$\sact=_{b'} \sact'$, $\sl\otu,\g\f\sr\rto{b_2, {\sact'}} \fdnu$, and $(\g\e\bullet \fdmu) \s^{b'} (\g\f\bullet \fdnu)$.
\end{enumerate}
\end{definition}

Two configurations $\sl \ott, \e\sr$ and $\sl \otu, \f\sr$ are symbolically $b$-bisimilar, denoted by  $\sl \ott, \e\sr\bis^b \sl \otu, \f\sr$, if there exists a symbolic bisimulation $\S=\{\s^b : b\in BExp\}$ such that $\osnt\s^b\osnu$.
Two quantum process terms $\ott$ and $\otu$ are symbolically $b$-bisimilar, denoted by $\ott\bis^b \otu$, if $\sl \ott, \id_\h\sr\bis^{b} \sl \otu, \id_\h\sr$. When $b=\true$, we simply write $\ott\bis\otu$.

To show the usage of symbolic bisimulation, we revisit the examples presented in Section~\ref{subsec:sts} to show that the proposed protocols indeed achieve the desired goals.
Let $\widetilde{A}=\{A_i : i\in I\}$ be a set of disjoint subsets of snapshots. An equivalence relation $\s$ is said to be generated by $\widetilde{A}$ if its equivalence classes on the set of snapshots $\cup_{i\in I}A_i$ 
are given by the partition $\widetilde{A}$, and it is the identity relation on $SN-\cup_{i\in I}A_i$.

\begin{example}\rm (Example~\ref{exm:set0} revisited)  This example is devoted
to showing rigorously that the two ways of setting a quantum system to the pure state $|0\>$,
presented in Examples~\ref{exm:set0c} and~\ref{exm:set0}, are indeed bisimilar. Let 
\begin{eqnarray*}
A&=&\{\sl\ctp, \id_\h \sr, \sl \ctq, \id_\h \sr\},\\
B &=&\{\sl \id[q].\nil, Set^0_q\sr, \sl \ctq_0, Set^0_q\sr, \sl \ctq_1, Set^1_q\sr\}
\end{eqnarray*}
and $\s'$ be the equivalence relation generated by $\{A, B\}$. It is easy to check that the family
$\{\s^b : b\in BExp\}$, where $\s^b=\s'$ for any $b\in BExp$, is a symbolic bisimulation. Thus $P\bis Q$.
\end{example}

\begin{example}\rm  (Superdense coding revisited) This example is devoted
to proving rigorously that the protocol presented in Example~\ref{exm:sdc}
indeed sends two bits of classical information from Alice to
Bob by transmitting a qubit. For that purpose, we need to show that $\sl Sdc_{spec}, \id_\h\sr\bis^{\true} \sl Sdc, \id_\h\sr$. Indeed, let 
\begin{eqnarray*}
A&=&\{\sl Sdc_{spec}, \id_\h\sr, \sl Sdc, \id_\h\sr\},\\
B^j&=&\{\osnt :  d(\osnt)=j\},\\
C^k_i&=&\{\osnt : \osnt \mbox{ along the branch of $x=i$, and } d(\osnt)=k\},
\end{eqnarray*}
where $d(\osnt)$ is the depth of the node $\osnt$ from the root of its corresponding qLTS, $0< j\leq 4$, $0\leq i\leq 3$, and $5\leq k\leq 10$. Let $\s_1^{\true}$ be the equivalence relation generated by $\{A, B^1, B^2, B^3, B^4\}$, and $\s_1^{x=i}$ generated by $\{C^k_i : 5\leq k\leq 10\}$.
For any $b\in BExp$, let $\s^b$ be $\s_1^{x=i}$ if $b\leftrightarrow x=i$, $\s_1^{\true}$ if $b\leftrightarrow \true$, and $\qc{Id}_{SN}$ otherwise.
Then it is easy to check that $\S=\{\s^{b} : b\in BExp\}$ is a symbolic bisimulation.
\end{example}

In the following, we denote by $\s^*$ the equivalence closure of a relation $\s$.

\begin{definition}
A relation family $\S=\{\s^b : b\in BExp\}$ is called decreasing, if for any $b,b'\in BExp$ with $b\ra b'$, we have $\s^{b'}\subseteq \s^{b}$. 
\end{definition}

\begin{lemma}\label{lem:makedec}
Let $\S=\{\s^b : b\in BExp\}$ be a symbolic bisimulation. Then there exists a decreasing symbolic bisimulation $\U=\{\u^{b} : b\in BExp\}$ such that for each $b\in BExp$, $\s^b\subseteq \u^{b}$.
\end{lemma}
{\it Proof.} Suppose $\S=\{\s^b : b\in BExp\}$ is a symbolic bisimulation. For each $b\in BExp$, let
 $$\u_1^{b}=\bigcup\{\s^{b'} : b\ra b'\}\mbox{  and  } \u^{b}=(\u_1^{b})^*.$$
Obviously, $\U=\{\u^{b} : b\in BExp\}$ is decreasing. 
We have to show that $\U$ is a symbolic bisimulation.

Let $b\in BExp$ and $\osnt \u^b\osnu$. Note that $\u_1^b$ is both reflexive and symmetric. So $\u^b$ is actually the transitive closure of $\u_1^b$, and there exist $n\geq 1$ and a sequence of snapshots $\sl  \ott_i,\e_i\sr$, $0\leq i\leq n$, such that $\osnt=\sl  \ott_0,\e_0\sr$, $\sl  \otu,\f\sr=\sl  \ott_n,\e_n\sr$, and for each $0\leq i\leq n-1$, $\sl  \ott_i,\e_i\sr  \u_1^b \sl  \ott_{i+1},\e_{i+1}\sr$. For the sake of simplicity, we assume $n=2$. That is, there exists $\osns$ such that $\osnt \s^{b_1}\osns\s^{b_2}\osnu$ with $b\ra b_1\wedge b_2$. The general case is more tedious but similar.

First we check that if $b$ is satisfiable, then
$qv(\ott)=qv(\ots)=qv(\otu)$ and $\e\eqsim_{\overline{qv(\ott)}} \g\eqsim_{\overline{qv(\ott)}} \f$. Now for any $\g'\in CP_t(\h_{\overline{qv(\ott)}})$, suppose $\sl \ott,\g'\e\sr \rto{b'_1, \sact} \fdmu$ with $bv(\sact)\cap fv(b_1, \ott, \otu)=\emptyset$. By $\alpha$-conversion, we may assume further that $bv(\sact)\cap fv(\ots)=\emptyset$. From $\osnt \s^{b_1}\osns$, there exists a collection of booleans $\{c_i : 1\leq i\leq n\}$ such that $b_1\wedge b_1'\ra \bigvee  c_i$ and  
for any $i$, $\exists c_i', \sact_i$ with $c_i\ra c_i'$,
$\sact=_{c_i} \sact_i$, $\sl \ots,\g'\g\sr\rto{c_i', {\sact_i}} \fdomega$, and $(\g'\e\bullet \fdmu) \s^{c_i} (\g'\g\bullet \fdomega)$. 
By $\alpha$-conversion, we can again assume that for each $i$, $bv(\sact_i)\cap fv(b_2, \ots, \otu)=\emptyset$.
Now by the assumption that $\osns\s^{b_2}\osnu$, there exists a 
collection of booleans $\{d_{ij} : 1\leq j\leq n_i\}$ such that
$b_2\wedge c_i'\ra \bigvee_j  d_{ij}$ and for any $d_{ij}$, $\exists d'_{ij}, \sact_{ij}$ with $d_{ij}\ra d'_{ij}$,
$\sact_{ij}=_{d_{ij}} \sact_i$, $\sl \otu,\g'\f\sr\rto{d'_{ij}, {\sact_{ij}}} \fdnu$, and $(\g'\g\bullet \fdomega) \s^{d_{ij}} (\g'\f\bullet \fdnu)$. 

Now let $$B=\{b\wedge c_i\wedge d_{ij} : 1\leq i\leq n, 1\leq j\leq n_i\}.$$
From the fact that $b\ra b_1\wedge b_2$, it is easy to check that $b\wedge b'_1\ra \bigvee B$. For any $c=b\wedge c_i\wedge d_{ij}$, we take $c'= d'_{ij}$ and $\sact'=\sact_{ij}$. Then $c\ra c'$, $\sact'=_c \sact$, and $\sl \otu,\g'\f\sr\rto{c', {\sact'}} \fdnu$. Furthermore, by 
the fact that $c\ra c_i$ and the definition of $\u^c$, we have $(\g'\e\bullet \fdmu) \u^{c} (\g'\g\bullet \fdomega)$ indeed. Similarly, $(\g'\g\bullet \fdomega) \u^{c} (\g'\f\bullet \fdnu)$. Thus  $(\g'\e\bullet \fdmu) \u^{c} (\g'\f\bullet \fdnu)$ as required. \hfill $\Box$

\begin{lemma}\label{lem:09272} Let decreasing families $\S_i=\{\s_i^b : b\in BExp\}$, $i=1,2$, be symbolic bisimulations. Then the family
$\S=\{(\s_1^b\s_2^b)^* : b\in BExp\}$ is also a symbolic bisimulation.
\end{lemma}
{\it Proof.} Let $b\in BExp$ and $\osnt (\s_1^b\s_2^b)^*\osnu$. Suppose there exist $n\geq 1$ and a sequence of snapshots $\sl  \ott_i,\e_i\sr$, $0\leq i\leq n$, such that $\osnt=\sl  \ott_0,\e_0\sr$, $\sl  \otu,\f\sr=\sl  \ott_n,\e_n\sr$, and for each $0\leq i\leq n-1$, $\sl  \ott_i,\e_i\sr  \s_1^b\s_2^b \sl  \ott_{i+1},\e_{i+1}\sr$. Again, for the sake of simplicity, we assume $n=1$. That is, there exists $\osns$ such that $\osnt \s_1^b\osns\s_2^b\osnu$. The rest of the poof follows almost the same lines of those in Lemma~\ref{lem:makedec}, by employing the assumption that  $\S_1$ and $\S_2$ are both decreasing. \hfill $\Box$

\begin{lemma}
Let $\S=\{\s^b : b\in BExp\}$ be a symbolic bisimulation and $c\in BExp$. Then $\S_c=\{\u^{b} = \s^{b\vee c} : b\in BExp\}$ is also a symbolic bisimulation.
\end{lemma}
{\it Proof.} Easy from definition.\hfill $\Box$

\begin{corollary}\label{cor:appupc}
If $b\ra b'$, then $\bis^{b'}{\subseteq} \bis^b$. That is, the relation family $\{\bis^b : b\in BExp\}$ is decreasing.
\end{corollary}

With the lemmas above, we can show that the family $\{\bis^b : b\in BExp\}$ is actually the largest symbolic bisimulation.
\begin{theorem}
\begin{enumerate}
\item For each $b\in BExp$, $\bis^b$ is an equivalence relation.
\item The family $\{\bis^b : b\in BExp\}$ is a symbolic bisimulation.
\end{enumerate}
\end{theorem}
{\it Proof.} (2) is direct from (1). To prove (1), let $b\in BExp$. Obviously, $\bis^b$ is reflexive and symmetric. To show the transitivity of $\bis^b$, let $\osnt\bis^b\osnu$ and
$\osnu\bis^b\osns$. Then by definition, there exist symbolic bisimulations $\S_i=\{\s_i^b : b\in BExp\}$, $i=1,2$, such that $\osnt\s_1^b\osnu$ and $\osnu\s_2^b\osns$. By Lemma~\ref{lem:makedec}, we can assume without loss of generality that both $\S_1$ and $\S_2$ are decreasing, thus $\S=\{(\s_1^b\s_2^b)^* : b\in BExp\}$ is also a symbolic bisimulation, by Lemma~\ref{lem:09272}.
So $\osnt\bis^b\osns$.  \hfill $\Box$

To conclude this subsection, we present a property of symbolic bisimilarity which is useful for the next section.

\begin{theorem}\label{theorem:sbisimulation}
Let $\osnt, \osnu\in SN$ and $b\in BExp$. Then $\osnt\bis^b \osnu$ if and only if
\begin{enumerate}
\item
$qv(\ott)=qv(\otu)$ and $\e\eqsim_{\overline{qv(\ott)}} \f$, if $b$ is satisfiable;
\item for any $\g\in CP_t(\h_{\overline{qv(\ott)}})$, whenever $\sl\ott,\g\e\sr \rto{b_1, \sact} \fdmu$ with $bv(\sact)\cap fv(b, \ott, \otu)=\emptyset$, then there exist a collection of booleans $B$ such that $b\wedge b_1\ra\bigvee B$ and $\forall\ b'\in B$, $\exists b_2, \sact'$ with $b'\ra b_2$,
$\sact=_{b'} \sact'$, $\sl\otu,\g\f\sr\rto{b_2, {\sact'}} \fdnu$, and $(\g\e\bullet \fdmu) \bis^{b'} (\g\f\bullet \fdnu)$;
\item Symmetric condition of (2).
\end{enumerate}
\end{theorem}
{\it Proof.} Routine. \hfill $\Box$

\subsection{Connection of symbolic and open bisimulations}
To ease notation, in the rest of the paper we use $\st, \su$ to range over $SN$, and 
sometimes equate $\st$ with $\osnt$, $\su$ with $\osnu$, $\fdmu$ with $\sum_{i\in I}\a_i\bullet \sl \ott_i, \e_i\sr$, and 
$\fdnu$ with $\sum_{j\in J}\b_j\bullet \sl \otu_j, \f_j\sr$ without stating them explicitly. We also write
 $$(\fdmu\eval)(\rho)= \sum_{i\in I}\tr(\a_i(\rho)) \< \ott_i\eval, \e_i(\rho)\>\mbox{  and  } (\fdnu\eval)(\rho)= \sum_{j\in J}\tr(\b_j(\rho)) \< \otu_j\eval, \f_j(\rho)\>.$$ 
In particular, $(\st\eval)(\rho)=\<\ott\eval,\e(\rho)\>$ and $(\su\eval)(\rho)=\<\otu\eval,\f(\rho)\>$.
The basic ideas of the proofs in this subsection are borrowed from~\cite{HL95}, with the help of Lemma~\ref{lem:so2s} and~\ref{lem:ss2o}.

Let $\S=\{\s^b : b\in BExp\}$ be a symbolic bisimulation. Define
$$\r_\S = \{((\st\eval)(\rho), (\su\eval)(\rho)) : \rho\in \dh \mbox{ and } \exists b, \eval(b)=\true\mbox{ and } \st\s^b \su \}.$$
We prove that $\r_\S$ is an open bisimulation. To achieve this, the following lemma is needed.

\begin{lemma}\label{lem:diss2o}
Let $\S=\{\s^b : b\in BExp\}$ be a symbolic bisimulation, $\rho\in \dh$, and $\eval(b)=\true$. Then
$$\fdmu\hspace{.2em} \s^b\hspace{.2em} \fdnu\mbox{ implies }(\fdmu\eval)(\rho)\hspace{.2em} \r_\S\hspace{.2em} (\fdnu\eval)(\rho).$$
\end{lemma}
{\it Proof.} Suppose $\fdmu = \sum_{i\in I}\a_i\bullet \sl \ott_i, \e_i\sr$,  
$\fdnu = \sum_{j\in J}\b_j\bullet \sl \otu_j, \f_j\sr$ and
 $\fdmu\hspace{.2em} \s^b\hspace{.2em} \fdnu$. We decompose the set $\supp\fdmu\cup\supp\fdnu$ into disjoint subsets $S_1, \cdots, S_n$ such that any two snapshots are in the same $S_k$ if and only if they are related by $\s^b$. 
For each $1\leq k\leq n$,
let 
$$K_{k}=\{i\in I : \sl \ott_i, \e_i\sr\in S_k\}\cup \{j\in J : \sl \otu_j, \f_j\sr\in S_k\}.$$ 
Then 
\begin{equation}\label{eq:09271}
\sum_{i\in K_{k}\cap I}\a_i\eqsim \sum_{j\in K_k\cap J}\b_j.
\end{equation}

For any $\rho\in \dh$ and $\eval$ such that $\eval(b)=\true$, 
\begin{eqnarray*}
(\fdmu\eval)(\rho)&=& \sum_{i\in I}\tr(\a_i(\rho)) \< \ott_i\eval, \e_i(\rho)\> = \sum_{k=1}^n \sum_{i\in K_k\cap I} \tr(\a_i(\rho))\< \ott_i\eval, \e_i(\rho)\> \\
&=&\sum_{k=1}^n \frac{1}{\sum_{j\in K_k\cap J} \tr(\b_j(\rho))}\sum_{i\in K_k\cap I}\sum_{j\in K_k\cap J} \tr(\a_i(\rho))\tr(\b_j(\rho))\< \ott_i\eval, \e_i(\rho)\>.
\end{eqnarray*}
Similarly, we have
\begin{eqnarray*}
(\fdnu\eval)(\rho)&=& \sum_{j\in J}\tr(\b_j(\rho)) \< \otu_j\eval, \f_j(\rho)\> = \sum_{k=1}^n \sum_{j\in K_k\cap J} \tr(\b_j(\rho))\< \otu_j\eval, \f_j(\rho)\> \\
&=&\sum_{k=1}^n \frac{1}{\sum_{i\in K_k\cap I} \tr(\a_i(\rho))}\sum_{i\in K_k\cap I}\sum_{j\in K_k\cap J} \tr(\a_i(\rho))\tr(\b_j(\rho))\< \otu_j\eval, \f_j(\rho)\>. 
\end{eqnarray*}
Note that by definition, if $\st\s^b \su$ then $(\st\eval)(\rho)\r_\S (\su\eval)(\rho)$. It follows that for any 
$1\leq k\leq n$, $i\in K_k\cap I$, and $j\in K_k\cap J$, we have $\< \ott_i\eval, \e_i(\rho)\> \r_\S \< \otu_j\eval, \f_j(\rho)\>$. Furthermore,
by Eq.(\ref{eq:09271}), we know $\sum_{i\in K_k\cap I} \tr(\a_i(\rho))=\sum_{j\in K_k\cap J} \tr(\b_j(\rho))$.
Thus $(\fdmu\eval)(\rho)\hspace{.2em} \r_\S\hspace{.2em} (\fdnu\eval)(\rho)$ by definition.\hfill $\Box$

\begin{lemma}\label{lem:rs} Let $\S=\{\s^b : b\in BExp\}$ be a symbolic bisimulation.
Then $\r_\S$ is an open bisimulation.
\end{lemma}
{\it Proof.} Let $(\st\eval)(\rho)\r_\S (\su\eval)(\rho)$. Then there exists $b$, such that $\eval(b)=\true$ and $\st\s^b \su.$ 
Thus we have 
\begin{enumerate}
\item $qv(\ott\eval)=qv(\ott)=qv(\otu)=qv(\otu\eval)$, and $\tr_{qv(\ott\eval)}\e(\rho) = \tr_{qv(\ott\eval)}\f(\rho)$ from $\e\eqsim_{\overline{qv(t)}} \f$. 
\item For any $\g\in CP_t(\h_{\overline{qv(\ott\eval)}})$, let 
$$\< \ott\eval, \g\e(\rho)\>\srto{\alpha} \dmu.$$ 
Then by Lemma~\ref{lem:so2s}, we have
$$\sl\ott,\g\e\sr\rto{b_1, \sact} \fdmu'=\sum_{i\in I} \a_i\bullet \sl \ott_i, \e_i\g\e\sr$$ 
such that $\eval(b_1)=\true$, 
$$\dmu=\sum_{i\in I}\tr(\a_i\g\e(\rho))\<\ott_i\eval', \e_i\g\e(\rho)\>.$$
Furthermore, we have $\sact = c?x$ for some $x\not\in fv(t)$ and $\eval'=\eval\{v/x\}$ if $\alpha=c?v$, or $\sact=_\eval\alpha$ and $\eval'=\eval$ otherwise.
Note that if $\sact=c?x$, we can always take $x$ such that $x\not\in fv(t,u,b)$ by $\alpha$-conversion.
Now by the assumption that  $\st\s^b \su$,  there exists a collection of booleans $B$ such that $b\wedge b_1\ra \bigvee B$ and $\forall\ b'\in B$, $\exists b_2, \sact'$ with $b'\ra b_2$,
$\sact=_{b'}\sact'$, 
$$\sl\otu,\g\f\sr\rto{b_2, {\sact'}} \fdnu'=\sum_{j\in J}\b_j\bullet \sl \otu_j, \f_j\g\f\sr,$$ 
and $(\g\e\bullet\fdmu')\s^{b'} (\g\f\bullet\fdnu')$. Note that $\eval(b\wedge b_1)=\true$ and $b\wedge b_1\ra \bigvee B$. We can always find a $b'\in B$ such that
$\eval(b')=\true$, and so $\eval(b_2)=\true$ as well. Then by Lemma~\ref{lem:ss2o}, we have 
$$\< \otu\eval, \g\f(\rho)\>\srto{{\beta}} \dnu=\sum_{j\in J}\tr(\b_j\g\f(\rho))\<\otu_j\eval'', \f_j\g\f(\rho)\>$$
where $\beta = c?v$ and $\eval''=\eval\{v/x\}$ if $\sact'=c?x$, or $\sact'=_\eval\beta$ and $\eval''=\eval$ otherwise.

We claim that $\beta = \alpha$, and $\eval'' = \eval'$. There are three cases to consider:
\begin{enumerate}
\item[(i)] $\alpha = c?v$. Then $\sact = c?x$ and $\eval'=\eval\{v/x\}$. So $\sact'= c?x$ by definition, which implies that $\beta=c?v=\alpha$, and $\eval''=\eval\{v/x\}=\eval'$.
\item[(ii)] $\alpha = c!v$. Then $\sact = c!e$, $\eval(e) = v$, and $\eval'=\eval$. So $\sact'= c!e'$ with $b'\ra e=e'$, which implies that $\beta=c!v'$ where $v'=\eval(e')$, and $\eval''=\eval=\eval'$. Finally, from $\eval(b')=\true$ we deduce $v'=v$.
\item[(iii)] For other cases, $\beta=\sact'=\sact=\alpha$, and $\eval''=\eval=\eval'$.
\end{enumerate}
Finally, by Lemma~\ref{lem:diss2o} we deduce $\mu\r_\S \nu$ from the facts that $(\g\e\bullet\fdmu')\s^{b'} (\g\f\bullet\fdnu')$ and $\eval'(b')=\true$.\hfill $\Box$
\end{enumerate}

\begin{corollary}
Let $b\in BExp$, $\ott, \otu\in \t$, and $\ctp, \ctq\in \p$. Then
\begin{enumerate} 
\item
$\ott\bis^b \otu$ implies for any evaluation $\eval$, if $\eval(b)=\true$ then $\ott\eval\stbis\otu\eval$. 
\item  $\ott\bis \otu$ implies $\ott\stbis\otu$. 
\item  $\ctp\bis^b \ctq$ implies $\ctp\stbis  \ctq$, provided that $b$ is satisfiable.
\end{enumerate}
\end{corollary}
{\it Proof.} (2) and (3) are both direct corollaries of (1). To prove (1), let $\ott\bis^b\otu$, and $\S=\{\s^b : b\in BExp\}$ be a symbolic bisimulation such that $\sl\ott, \id_\h\sr\s^b\sl\otu, \id_\h\sr$. Then by Lemma~\ref{lem:rs}, for any evaluation $\eval$ and any $\rho$, $\eval(b)=\true$ implies $\< \ott\eval, \rho\>\stbis \< \otu\eval, \rho\>$. Thus $\ott\eval\stbis \otu\eval$ by definition.  \hfill $\Box$

For any $b\in BExp$, define
$$\s_{\stbis}^b = \{(\st, \su): \forall \eval, \eval(b)=\true \mbox{ implies that for any } \rho\in\dh, (\st\eval)(\rho)\stbis(\su\eval)(\rho)\}.$$ 
We prove that $\S_{\stbis}=\{\s_{\stbis}^b : b\in BExp\}$ is a symbolic bisimulation. Firstly, it is easy to check that for each $b$, $\s_{\stbis}^b$ is an equivalence relation.
Two quantum states $\rho, \sigma\in \dh$ are said to be $equal$ $except$ $at$ $\widetilde{q}$ if $\tr_{\widetilde{q}} \rho = \tr_{\widetilde{q}} \sigma$.
Then we can show the following lemma, which is parallel to Lemma~\ref{lem:diss2o}.

\begin{lemma}\label{lem:diso2s} Let $b\in BExp$.
If for any evaluation $\eval$,
$$\eval(b)=\true\mbox{ implies that }\forall \rho\in \dh, (\fdmu\eval)(\rho)\hspace{.2em} \stbis\hspace{.2em} (\fdnu\eval)(\rho),$$
then $\fdmu\hspace{.2em} \s_{\stbis}^b\hspace{.2em}\fdnu.$
\end{lemma}
{\it Proof.} Let
 $\fdmu = \sum_{i\in I}\a_i\bullet \sl \ott_i, \e_i\sr$ and
$\fdnu = \sum_{j\in J}\b_j\bullet \sl \otu_j, \f_j\sr$.
We prove this lemma by distinguishing two cases:
\begin{enumerate}
\item Both $|I|>1$ and $|J|>1$. Similar to Lemma~\ref{lem:diss2o}, we first decompose the set $\supp\fdmu\cup\supp\fdnu$ into disjoint subsets $S_1, \cdots, S_n$ such that any two snapshots are in the same $S_k$ if and only if they are related by $\s_{\stbis}^b$. 
For each $1\leq k\leq n$,
let 
\begin{equation}\label{eq:partition}
K_{k}=\{i\in I : \sl \ott_i, \e_i\sr\in S_k\}\cup \{j\in J : \sl \otu_j, \f_j\sr\in S_k\}
\end{equation}
and $\K=\{K_k: 1\leq k\leq n\}$. Note that by Lemma~\ref{lem:superoperatorc}, there are two sets of pairwise orthogonal pure states $\{|\phi_i\> :  i\in I\}$ and $\{|\phi'_j\> : j\in J\}$ in some $\h_{\widetilde{q}}$ such that 
the Kraus operators of $\a_i$ and $\e_i$ are $\{|\phi_i\>\<\phi_i|\}$ and $\{|\phi_i\>\<\phi_{i'}| : i'\in I\}$, respectively, while the Kraus operators of $\b_j$ and $\f_j$ are $\{|\phi'_j\>\<\phi'_j|\}$ and $\{|\phi'_j\>\<\phi'_{j'}| : j'\in J\}$, respectively.  
Let $E_k=\sum_{i\in K_{k}\cap I}|\phi_i\>\<\phi_i|$, and $F_k=\sum_{j\in K_{k}\cap J}|\phi_j'\>\<\phi_j'|$.
Then it suffices to show $E_k= F_k$, $1\leq k\leq n$. In the following, we prove $E_1=F_1$; other cases are similar.

For any $\rho$ and $\eval$ such that $\eval(b)=\true$, we decompose the set $\supp{(\fdmu\eval)(\rho)}\cup\supp{(\fdnu\eval)(\rho)}$ into equivalence classes 
$R_1, \cdots, R_{m^\eval_\rho}$ according to $\stbis$. For each $1\leq l\leq {m^\eval_\rho}$, let  
$$L_{l}^{\eval, \rho}=\{i\in I : \< \ott_i\eval, \e_i(\rho)\>\in R_l\}\cup \{j\in J : \< \otu_j\eval, \f_j(\rho)\>\in R_l\}
$$
and $\L^{\eval, \rho} = \{L_{l}^{\eval, \rho}: 1\leq l\leq R_{m^\eval_\rho}\}$.
Note that by definition, $\K$ is a refinement of $\L^{\eval, \rho}$ for any $\eval(b)=\true$ and $\rho$. We assume without loss of generality that $L_1^{\eval, \rho}$ is the partition in $\L^{\eval, \rho}$ which contains $K_1$, and $L_1^{\eval, \rho}=K_1 \cup K_1^{\eval, \rho}$  where $K_1^{\eval, \rho}=\bigcup_{k\in I_{\eval, \rho}} K_k$, $I_{\eval, \rho}$ is a subset of $\{2,\cdots, n\}$.

As the effects of the super-operators $\a_i$ and $\b_j$ are simply erasing the original information at $\widetilde{q}$ and setting the partial states of $\widetilde{q}$ to be $|\phi_i\>$ and $|\phi_j'\>$, respectively,  we have
$\L^{\eval, \rho}=\L^{\eval, \sigma}$ (which means $m^\eval_\rho=m^\eval_\sigma$, and $L_l^{\eval, \rho}=L_l^{\eval, \sigma}$ for each $l$) for all $\sigma$ which is equal to $\rho$ except at $\widetilde{q}$. Note that $\tr(\a_i(\rho))=\tr(|\phi_i\>_{\widetilde{q}}\<\phi_i|\rho)=\tr(|\phi_i\>_{\widetilde{q}}\<\phi_i|\rho_{\widetilde{q}})$ where $\rho_{\widetilde{q}} = \tr_{\overline{\widetilde{q}}} \rho$ is the reduced state of $\rho$ at the systems $\widetilde{q}$. 
Let $E_1^{\eval, \rho}=\sum_{k\in I_{\eval, \rho}} E_k$ and $F_1^{\eval, \rho}=\sum_{k\in I_{\eval, \rho}} F_k$.
Then 
for any $\rho'\in \d(\h_{\widetilde{q}})$,
$$\tr((E_1 + E_1^{\eval, \rho})\rho')=\sum_{i\in L_{1}^{\eval, \sigma}\cap I}\tr(\a_i(\sigma))=\sum_{j\in L_{1}^{\eval, \sigma}\cap J}\tr(\b_j(\sigma))=\tr((F_1 +F_1^{\eval, \rho})\rho')$$
where $\sigma=\rho'\otimes \tr_{\widetilde{q}}(\rho)$ is equal to $\rho$ except at $\widetilde{q}$, and the second equality is from the assumption that $(\fdmu\eval)(\sigma)\stbis(\fdnu\eval)(\sigma)$. This implies $E_1 + E_1^{\eval, \rho}=F_1 +F_1^{\eval, \rho}$. 

Let $K=\bigcap_{\rho, \eval(b)=\true}  I_{\eval, \rho}$. We claim that $K=\emptyset$. Otherwise, there exists $k$ such that $k\in I_{\eval, \rho}$ for any $\eval(b)=\true$ and $\rho$. Then by the definition of $L_1^{\eval, \rho}$, we have $\< \ott_i\eval, \e_i(\rho)\>\stbis \< \ott_{i'}\eval, \e_{i'}(\rho)\> $ where $i\in K_1$ and $i'\in K_k$. Thus $\sl \ott_i, \e_i\sr \s_{\stbis}^b \sl\ott_{i'}, \e_{i'}\sr$, contradicting the fact that they belong to different equivalence classes of $\s_{\stbis}^b$. 

Now for any pure state $|\phi\>$ such that $E_1|\phi\>=|\phi\>$,
we have
 $E_1^{\eval, \rho}|\phi\>=0$ for any $\rho$ and $\eval(b)=\true$, by the orthogonality of $E_i$'s. Thus $F_1^{\eval, \rho}|\phi\>=|\phi\>-F_1|\phi\>$.
Note that $F_1^{\eval', \rho'}F_1^{\eval, \rho}=\sum_{k\in I_{\eval, \rho}\cap I_{\eval', \rho'}} F_k=F_1^{\eval, \rho}F_1^{\eval', \rho'}$.
We have 
$$\sum_{k\in I_{\eval, \rho}\cap I_{\eval', \rho'}} F_k|\phi\>=|\phi\>-F_1|\phi\>,$$
and finally, $\sum_{k\in K} F_k|\phi\>=|\phi\>-F_1|\phi\>.$ Then $F_1|\phi\>=|\phi\>$ from the fact that $K=\emptyset$. Similarly, we can prove that
for any $|\phi\>$, $F_1|\phi\>=|\phi\>$ implies $E_1|\phi\>=|\phi\>$. Thus $E_1=F_1$.

\item Either $|I|=1$ or $|J|=1$. Let us suppose $|I|=1$, and $\fdmu=\osnt$. We need to show that for each $j\in J$, $\b_j\neq 0_\h$ implies $\osnt \s_{\stbis}^b \sl\otu_j, \f_j\sr$. This is true because otherwise we can find $\eval(b)=\true$, $j\in J$, and $\rho\in \dh$ such that $\tr(\b_j(\rho))\neq 0$ but $\<\ott\eval, \e(\rho)\>\nssbis \<\otu_j\eval, \f_j(\rho)\>$. Thus 
$(\fdmu\eval)(\rho)\nssbis (\fdnu\eval)(\rho)$, a contradiction. \hfill $\Box$
\end{enumerate}

\begin{lemma}\label{lem:sr} The family $\S_{\stbis}=\{\s_{\stbis}^b : b\in BExp\}$ is a symbolic bisimulation.
\end{lemma}
{\it Proof.} Let $b\in BExp$ and $\st\s_{\stbis}^b \su$. Then for any $\eval$, $\eval(b)=\true$ implies that for any $\rho\in\dh$, $(\st\eval)(\rho)\stbis(\su\eval)(\rho)$. Thus we have 
\begin{enumerate}
\item  If $b$ is satisfiable, then $qv(\ott)=qv(\ott\eval)=qv(\otu\eval)=qv(\otu)$, and $\e\eqsim_{\overline{qv(\ott)}} \f$ from the fact that $\tr_{qv(\ott)}\e(\rho) = \tr_{qv(\ott)}\f(\rho)$ for any $\rho$. 
\item For any $\g\in CP_t(\h_{\overline{qv(\ott)}})$, let 
\begin{equation}\label{eq:tmp09251}
\sl\ott,\g\e\sr \rto{b_1, \sact} \fdmu'=\sum_{i\in I} \a_i\bullet \sl \ott_i, \e_i\g\e\sr
\end{equation}
 with $bv(\sact)\cap fv(b, \ott, \otu)=\emptyset$.
We need to construct a set of booleans $B$ such that $b\wedge b_1\ra\bigvee B$, and 
$\forall\ b'\in B$, $\exists b_2, \sact'$ with $b'\ra b_2$,
$\sact=_{b'} \sact'$, $\sl\otu,\g\f\sr\rto{b_2, {\sact'}} \fdnu'$, and $(\g\e\bullet \fdmu') \s^{b'}(\g\f\bullet \fdnu')$. Let $$U=\{\fdomega : \sl\otu,\g\f\sr \rto{b(\fdomega), \sact(\fdomega)}\fdomega \mbox{ and } \sact=_{\false}\sact(\fdomega)\}.$$ Here
similar to~\cite{HL95}, to ease the notations we only consider the case where for each $\fdomega$, there is at most one symbolic action, denoted by $(b(\fdomega), \sact(\fdomega))$, such that $\sl\otu,\g\f\sr \rto{b(\fdomega), \sact(\fdomega)}\fdomega$.
For each $\fdomega\in U$, let $b'_{\fdomega}$ be a boolean expression such that for any $\eval$,
\begin{equation}\label{eq:tmp0111}
\eval(b'_{\fdomega})=\true\mbox{   if and only if  for any }\rho,  (\g\e\bullet \fdmu'\tilde{\eval})(\rho) \stbis (\g\f\bullet\fdomega\tilde{\eval})(\rho)
\end{equation}
where $\tilde{\eval}=\eval\{v/x\}$ for some $v$ if $\sact=c?x$, and $\tilde{\eval}=\eval$ otherwise. 

Let $B=\{b_{\fdomega} : \fdomega\in U\}$, where $b_{\fdomega}=b'_{\fdomega}\wedge b''_{\fdomega} \wedge b(\fdomega)$ and $b''_{\fdomega}$ is a boolean expression defined by
\begin{equation}\label{eq:tmp010111}
b''_{\fdomega}\equiv\begin{cases}\ e = e' &
\mbox{if }\sact = c!e \mbox{ and }\sact(\fdomega) = c!e' \mbox{ are both classical output}, \\
\ \true  & \mbox{otherwise}.\end{cases}
\end{equation}
Then obviously, $\sact=_{b_\fdomega}\sact(\fdomega)$. We check $b\wedge b_1 \ra \bigvee B$. For any evaluation $\eval$ such that $\eval(b\wedge b_1)=\true$, we have by definition of $\s_{\stbis}^b$ that $\< \ott\eval, \e(\rho)\>\stbis\< \otu\eval, \f(\rho)\>$  for any $\rho$. On the other hand, by Lemma~\ref{lem:ss2o} and Eq.(\ref{eq:tmp09251}), we obtain
$$\< \ott\eval, \g\e(\rho)\>\srto{\alpha} \dmu=\sum_{i\in I}\tr(\a_i\g\e(\rho))\<\ott_i\eval', \e_i\g\e(\rho)\>$$
where $\alpha=c?v$ and $\eval'=\eval\{v/x\}$ if $\sact=c?x$, and $\alpha=_{\eval}\sact$ and $\eval'=\eval$ otherwise.
To match this transition, we have
$$\< \otu\eval, \g\f(\rho)\>\srto{{\alpha}} \dnu$$
for some $\dnu$ such that $\dmu\stbis \dnu$. 
Now from Lemma~\ref{lem:so2s}, there exists $\fdnu'\in U$  such that  $\eval(b(\fdnu'))=\true$,
$$\sl\otu,\g\f\sr\rto{b(\fdnu'), {\sact(\fdnu')}} \fdnu'=\sum_{j\in J}\b_j\bullet \sl \otu_j, \f_j\g\f\sr,$$ 
$$\dnu =\sum_{j\in J}\tr(\b_j\g\f(\rho))\<\otu_j\eval'', \f_j\g\f(\rho)\>.$$
Furthermore, we have $\sact(\fdnu')=c?y$ for some $y\not\in fv(u)$ and $\eval''=\eval\{v/y\}$ if $\alpha=c?v$, and
$\alpha=_{\eval}\sact(\fdnu')$ and $\eval''=\eval$ otherwise.

We claim that $\sact=_{\eval}\sact(\fdnu')$, and $\eval'' = \eval'$. There are three cases to consider:
\begin{enumerate}
\item[(i)] $\sact = c?x$. Then $\alpha = c?v$ and $\eval'=\eval\{v/x\}$, which implies that $\sact(\fdnu')=c?y$ for some $y\not\in fv(u)$.
By $\alpha$-conversion and the fact that $x\not\in fv(b, t ,u)$, we can also take $y=x$. So $\sact(\fdnu')=\sact$, and $\eval''=\eval\{v/x\}=\eval'$.
\item[(ii)] For other cases, $\sact(\fdnu')=_{\eval}\alpha=_{\eval}\sact$, and $\eval''=\eval=\eval'$.
\end{enumerate}
Now we have $\dmu=(\g\e\bullet \fdmu'\eval')(\rho)$ and  $\dnu=(\g\f\bullet \fdnu'\eval')(\rho)$. From the arbitrariness of $\rho$, we know $\eval(b'_{\fdnu'})=\true$ from Eq.(\ref{eq:tmp0111}). By Eq.(\ref{eq:tmp010111}) and the fact that $\sact=_{\eval}\sact(\fdnu')$, we further derive that $\eval(b''_{\fdnu'})=\true$. Therefore, $\eval(b_{\fdnu'})=\true$, and so $\eval(\bigvee B)=\true$.

For any $b_{\fdomega}\in B$, we have $b_{\fdomega}\ra b(\fdomega)$,
$\sact=_{b(\fdomega)}\sact(\fdomega)$, and $\sl\otu,\g\f\sr \rto{b(\fdomega), \sact(\fdomega)}\fdomega$ by definition of $B$. Finally, 
for any evaluation $\eval$, if $\eval(b_\fdomega)=\true$ then $\eval(b'_\fdomega)=\true$, and from Eq.(\ref{eq:tmp0111}) we have  $(\g\e\bullet \fdmu'\tilde{\eval})(\rho) \stbis (\g\f\bullet\fdomega\tilde{\eval})(\rho)$ for any $\rho\in \dh$. Then $(\g\e\bullet \fdmu') \s^{b_\fdomega}(\g\f\bullet \fdomega)$ follows by Lemma~\ref{lem:diso2s}.
Here we have used that fact that $x\not\in fv(b, \ott, \otu)$ implies $\ott\eval\{v/x\}=\ott\eval$ and $\otu\eval\{v/x\}=\ott\eval$.\hfill $\Box$

\end{enumerate}

\begin{lemma}\label{lem:otos}
If for any evaluation $\eval$, $\eval(b)=\true$ implies $\ott\eval\stbis\otu\eval$, then $\ott\bis^b \otu$.
\end{lemma}
{\it Proof.} For any $\rho\in\dh$ and any evaluation $\eval$ such that $\eval(b)=\true$, we first derive $\< \ott\eval, \rho\>\stbis\< \otu\eval, \rho\>$ from the assumption that $\ott\eval\stbis\  \otu\eval$. Then by Lemma~\ref{lem:sr}, we have $\sl\ott, \id_\h\sr\bis^b\sl\otu, \id_\h\sr$, and thus $\ott\bis^b \otu$ by definition.  \hfill $\Box$

From the above lemmas, we finally reach our main result in this section.
\begin{theorem}
Let $b\in BExp$, $\ott, \otu\in \t$, and $\ctp, \ctq\in \p$. Then
\begin{enumerate} 
\item
$\ott\bis^b \otu$ if and only if for any evaluation $\eval$, $\eval(b)=\true$ implies $\ott\eval\stbis\otu\eval$. 
\item  $\ott\bis \otu$ if and only if $\ott\stbis\otu$. 
\item  $\ctp\bis^b \ctq$ if and only if $\ctp\stbis  \ctq$, provided that $b$ is satisfiable.
\end{enumerate}
\end{theorem}

\section{An algorithm for symbolic ground bisimulation}

From Clause (2) of Definition~\ref{def:ssbisimulation}, to check whether two snapshots are symbolically bisimilar, we are forced to compare their behaviours under any super-operators. This is generally infeasible since all super-operators constitute a continuum, and it seems hopeless to design an algorithm which works for the most general case. In this section, we develop an efficient algorithm for a class of quantum process terms which covers all existing practical quantum communication protocols. 
To this end, we first define the notion of symbolic ground bisimulation which stems from~\cite{San96}.
\begin{definition}\label{def:gsbisimulation}
A family of equivalence relations $\{\s^b : b\in BExp\}$ is called a symbolic ground bisimulation if for any $b\in BExp$, 
$\osnt\s^b\osnu$ implies that 
\begin{enumerate}
\item
$qv(\ott)=qv(\otu)$, and $\e\eqsim_{\overline{qv(\ott)}} \f$, 
\item whenever $\osnt \rto{b_1, \sact} \fdmu$ with $bv(\sact)\cap fv(b, \ott, \otu)=\emptyset$, then there exists a collection of booleans $B$ such that $b\wedge b_1\ra \bigvee B$ and $\forall\ b'\in B$, $\exists b_2, \sact'$ with $b'\ra b_2$,
$\sact=_{b'} \sact'$, $\osnu\rto{b_2, {\sact'}} \fdnu$, and $(\e\bullet \fdmu) \s^{b'} (\f\bullet \fdnu)$.
\end{enumerate}
\end{definition}

Given two configurations $\osnt$ and $\osnu$, we write $\osnt\bis^b_g \osnu$ if there exists a symbolic ground bisimulation $\{\s^b : b\in BExp\}$ such that $\osnt\s^b\osnu$.
\begin{definition}
A relation $\s$ on $SN$ is said to be closed under super-operator application if $\osnt\s\osnu$ implies $\sl\ott, \g\e\sr\s\sl\otu,\g\f\sr$ for any $\g\in CP_t(\h_{\overline{qv(\ott)}})$. A family of relations are closed under super-operator application if each individual relation is.\end{definition}

The following proposition, showing the difference of symbolic bisimulation and symbolic ground bisimulation, is easy from definition.
\begin{prop}\label{prop:biscg}
$\bis$ is the largest symbolic ground bisimulation that is closed under super-operator application.
\end{prop}

A process term is said to be {\it free of quantum input} if all of its descendants, including itself, can not perform quantum input actions.
Note that all existing quantum communication protocols such as super-dense coding~\cite{BW92}, teleportation~\cite{BB93}, quantum key-distribution protocols~\cite{BB84}, etc, are, or can easily modified to be, free of quantum input. Putting this constraint will not bring too much restriction on the application range of our algorithm.
\begin{lemma}\label{lem:freeqi}
Let $\osnt\bis^b_g\osnu$, and $\ott$ and $\otu$ both free of quantum input. Then for any $\g\in CP_t(\h_{\overline{qv(\ott)}})$, 
$\sl\ott, \g\e\sr\bis^b_g\sl\otu,\g\f\sr$.
\end{lemma}
{\it Proof.}
We need to show $\S=\{\s^b : b\in BExp\}$, where
$$\s^b = \{(\sl\ott,\g\e\sr, \sl\otu,\g\f\sr): \mbox{$\ott$ and $\otu$ free of quantum input, $\g\in CP_t(\h_{\overline{qv(\ott)}})$, and } \osnt\bis^b_g\osnu\},$$
is a symbolic ground bisimulation. This is easy by noting that for any descendant $\ott'$ of $\ott$, $qv(\ott')\subseteq qv(\ott)$, and then $\g\in CP_t(\h_{\overline{qv(\ott')}})$ as well. Consequently, $\g$ commutes with all the super-operators performed by $\ott$ and its descendants.
\hfill $\Box$

\begin{theorem}\label{thm:freeqi}
If $\ott$ and $\otu$ are both free of quantum input, then $\osnt\bis^b\osnu$ if and only if $\osnt\bis^b_g\osnu$.
\end{theorem}
{\it Proof.} Easy from Lemma~\ref{lem:freeqi}. \hfill $\Box$

Algorithm 1 computes the \emph{most general boolean} $b$
such that $\st\bis^b_g \su$, for two given snapshots $\st$ and $\su$. By the most
general boolean $mgb(\st,\su)$ we mean that $\st\bis_g^{mgb(\st,\su)} \su$ and
whenever $\st\bis_g^b \su$ then $b\rightarrow mgb(\st,\su)$. From Theorem~\ref{thm:freeqi}, this algorithm is applicable to 
verify the correctness of all existing quantum communication protocols.

The algorithm closely follows that introduced in \cite{HL95}. The main
procedure is $\textbf{Bisim}(\st,\su)$.  It starts with the initial
snapshot pairs $(\st,\su)$, trying to find the smallest symbolic bisimulation
relation containing the pair by comparing transitions from each pair
of snapshots it reaches.  The core procedure $\textbf{Match}$ has
four parameters: $\st$ and $\su$ are the current terms under examination;
$b$ is a boolean expression representing the constraints accumulated
by previous calls; $W$ is a set of snapshot pairs which have been
visited. For each possible action enabled by $\st$ and $\su$, the procedure $\textbf{MatchAction}$ is used to compare
possible moves from $\st$ and $\su$. Each comparison returns a boolean
and a table; the boolean turns out to be $mgb(\st,\su)$ and the table
is used to represent the witnessing bisimulation. We consider a table
as a function that maps a pair of snapshots to a boolean. The
disjoint union of tables, viewed as sets, is denoted by $\sqcup$.

\newsavebox{\tablebox}
\begin{algorithm}[t]
\begin{lrbox}{\tablebox}
{\small\caption{\textbf{Bisim}$(\st,\su)$}}
\parbox[c]{14cm}{
\begin{algorithmic}
\STATE \textbf{Bisim}$(\st,\su)=\textbf{Match}(\st,\su,\true,\emptyset)$ 

\bigskip
\textbf{Match}$(\st,\su,b,W)=$ \qquad\qquad where $\st=\sl\ott, \e\sr$ and $\su=\sl\otu, \f\sr$ 

\eIf{$(\st,\su)\in W$}{
$(\theta,T):=(\true,\emptyset)$
}
{ 
\For{$\sact\in Act(\st,\su)$}{
 $(\theta_\sact,T_\sact) := \textbf{MatchAction}(\sact,\st,\su,b,W)$
}
 $(\theta,T) := (\bigwedge_\sact \theta_\sact,\ 
\bigsqcup_\sact (T_\sact \sqcup \stet{(\st,\su)\mapsto (b\wedge \bigwedge_\sact \theta_\sact)}))$
}
\textbf{return} $(\theta\wedge (qv(\ott)=qv(\otu))  
\wedge (\e\eqsim_{\overline{qv(\ott)}} \f),T)$

\bigskip
\textbf{MatchAction}$(\sact,\st,\su,b,W)=$ 

\Switch{$\sact$}{
\Case{c!}{
\For{$\st\ar{b_i,c!e_i}\st_i$ and $\su\ar{b'_j,c!e'_j}\su_j$}{ 
$(\theta_{ij},T_{ij}) := \textbf{Match}(\st_i, \su_j, b\wedge b_i\wedge b'_j\wedge e_i=e'_j, \stet{(\st,\su)}\cup W)$ 
}
\STATE \textbf{return} $(\bigwedge_i(b_i\rightarrow \bigvee_j(b'_j\wedge e_i=e'_j\wedge \theta_{ij})) \wedge
\bigwedge_j(b'_j\rightarrow\bigvee_i(b_i\wedge e_i=e'_j\wedge \theta_{ij})),\ \bigsqcup_{ij}T_{ij})$
}

\Case{$\tau$}{
\For{$\st\ar{b_i,\tau}\Delta_i$ and $\su\ar{b'_j,\tau}\Theta_j$}{ 
$(\theta_{ij},T_{ij}) := \textbf{MatchDistribution}(\Delta_i,\Theta_j, b\wedge b_i\wedge b'_j, \stet{(\st,\su)}\cup W)$ 
}
\STATE \textbf{return} $(\bigwedge_i(b_i\rightarrow \bigvee_j(b'_j\wedge \theta_{ij})) \wedge
\bigwedge_j(b'_j\rightarrow\bigvee_i(b_i\wedge \theta_{ij})),\ \bigsqcup_{ij}T_{ij})$
}
\Other{\For{$\st\ar{b_i,\sact}\st_i$ and $\su\ar{b'_j,\sact}\su_j$} {
$(\theta_{ij},T_{ij}) := \textbf{Match}(\st_i, \su_j, b\wedge b_i\wedge b'_j, \stet{(\st,\su)}\cup W)$ 
} 
\textbf{return} $(\bigwedge_i(b_i\rightarrow \bigvee_j(b'_j\wedge \theta_{ij})) \wedge
\bigwedge_j(b'_j\rightarrow\bigvee_i(b_i\wedge \theta_{ij})),\ \bigsqcup_{ij}T_{ij})$
}
}

\bigskip
\textbf{MatchDistribution}($\Delta,\Theta, b, W$)=

\For{$\st_i\in\supp{\Delta}$ and $\su_j\in\supp{\Theta}$}{
$(\theta_{ij},T_{ij}) := \textbf{Match}(\st_i,\su_j,b,W)$
}
$\r := \stet{(\st,\su)\mid b \rightarrow (\bigsqcup_{ij} T_{ij})(\st,\su)}^*$

\textbf{return} (\textbf{Check}$(\Delta,\Theta,\r),\ \bigsqcup_{ij}T_{ij})$
\bigskip

\textbf{Check}$(\Delta,\Theta,\r)=$ 
 $\theta := \true$

\For{$S\in\ \supp{\Delta}\cup\supp{\Theta}/ \r$}{
$\theta := \theta\wedge (\Delta(S) \eqsim \Theta(S))$
}
\textbf{return} $\theta$ 
\end{algorithmic}
}
\end{lrbox}
\resizebox{.75\textwidth}{!}{\usebox{\tablebox}}
\end{algorithm}

The main difference from the algorithm of \cite{HL95} lies in the
comparison of $\tau$ transitions. We
introduce  the procedure $\textbf{MatchDistribution}$ to
approximate $\bis_g^b$ by a relation $\r$. 
For any two snapshots $\st_i\in\supp{\Delta}$ and $\su_j\in\supp{\Theta}$,
they are related by $\r$ if $b\rightarrow T(\st_i,\su_j)$. More
precisely, we use the equivalence closure of $\r$ instead in order for it to be used in the procedure
$\textbf{Check}$. Moreover, if a snapshot pair $(\st,\su)$ has been
visited before, i.e.  $(\st,\su)\in W$, then $T(\st,\su)$ is assumed
to be $\true$ in all future visits. Hence, $\r$ is coarser than
$\bis_g^b$ in general.
We use $\textbf{Check}(\Delta,\Theta,\r)$ to computate
the constraint so that the super-operator valued distribution $\Delta$ is related to
$\Theta$ by a relation lifted from $\r$. The correctness of the
algorithm is stated in the following theorem.

\begin{theorem}
For two snapshots $\st$ and $\su$, the function $\textbf{Bisim}(\st,\su)$ terminates. Moreover,
if $\textbf{Bisim}(\st,\su) = (\theta,T)$ then $T(\st,\su)=\theta=mgb(\st, \su)$.
\end{theorem}
{\it Proof.}
Termination is easy to show. Each time a new snapshot pair is
encountered, the procedure $\textbf{Match}$ is called and the pair is
added to the set $W$. Since we are considering a finitary transition
graph, the number of different pairs is finite.  Eventually every
possible pair is in $W$ and each call to $\textbf{Match}$ immediately
terminates.

Correctness of the algorithm is largely similar to that in \cite{HL95}, though we use the additional procedure \textbf{MatchDistribution} to compute the constraint that relates two super-operator valued distributions.
\hfill $\Box$

\section{Modal characterisation}

We now present a modal logic to characterise the behaviour of quantum snapshots and their distributions.

\begin{definition} The class $\l$ of quantum modal formulae over $Act_s$, ranged over by $\phi$, $\Phi$, etc, is defined by the following grammar:
\begin{eqnarray*}
\phi &::=& \g_{\widetilde{q}}\ |\ \neg \phi\ |\ \bigwedge_{i\in I} \phi_i\ |\ \g.\phi\ |\ \<\sact\>\Phi\\
\Phi &::=& Q_{\gtrsim \a}(\phi)\ |\ \bigwedge_{i\in I} \Phi_i
\end{eqnarray*}
where $\g\in CP_t(\h)$, $\sact\in Act_s$, and $\a\in CP(\h)$.
We call
$\phi$ a \emph{snapshot formula} and $\Phi$ a \emph{distribution
formula}. 
\end{definition}

The satisfaction relation $\models {\subseteq}\ {EV\times (SN\cup  \fdist(SN))\times \l}$ is defined as the minimal relation satisfying
\begin{itemize}
\item $\eval, \snt \models \g_{\widetilde{q}}$ if $qv(\ott)\cap \widetilde{q} = \emptyset$, and $\e\eqsim_{\widetilde{q}} \g$, where $\snt=\osnt$
;
\item $\eval, \snt \models \neg \phi$ if  $\eval, \snt \not\models \phi$;
\item $\eval, \snt \models  \bigwedge_{i\in I} \phi_i$ if $\eval, \snt  \models \phi_{i}$ for each $i\in I$;
\item $\eval, \snt \models  \g.\phi$ if $\g\in CP_t(\h_{\overline{qv(\ott)}})$ and $\sl\ott, \g\e\sr  \models \phi$, where $\snt=\osnt$;
\item $\eval, \snt  \models \<\sact\>\Phi$ if $\snt \rto{b, \sact'}\fdmu$ for some $b$, $\sact'$, and $\fdmu$, such that $\eval(b)=\true$, $\sact=_\eval \sact'$, and 
$\eval, \fdmu\models\Phi$;
\item $\eval, \fdmu\models Q_{\gtrsim \a}(\phi)$ if $$\sum_{\snt\in \supp \fdmu} \{\fdmu(\snt) : \eval, \snt\models \phi\} \gtrsim \a;$$
\item $\eval, \fdmu \models  \bigwedge_{i\in I} \Phi_i$ if $\eval, \fdmu  \models \Phi_{i}$ for each $i\in I$.
\end{itemize}

\begin{definition} Let $\eval$ be an evaluation. We write $\snt =^\eval_{\l}\snu$ if  for any $\phi\in \l$, $$\eval,\snt\models\phi \mbox{ if and only if } \eval, \snu\models\phi.$$ Similarly, $\fdmu =^\eval_{\l}\fdnu$ if  for any $\Phi\in \l$, $$\eval,\fdmu\models\Phi \mbox{ if and only if } \eval, \fdnu\models\Phi.$$
\end{definition}

\begin{lemma}\label{lem:witness} Let $\eval$ be an evaluation, $\snt, \snu\in SN$, and $\fdmu, \fdnu\in \fdist(SN)$.
\begin{enumerate}
\item If $\snt\not =^\eval_{\l}\snu$, then there exists $\phi\in \l$,  such that $\eval,\snt\models\phi$ but $\eval, \snu\not\models\phi;$
\item If $\fdmu\not =^\eval_{\l}\fdnu$, then there exists $\Phi\in \l$,  such that $\eval,\fdmu\models\Phi$ but $\eval, \fdnu\not\models\Phi.$
\end{enumerate}
\end{lemma}
{\it Proof.} (1) is easy as we have negation operator $\neg$ for state formulae. To prove (2), let $\fdmu\not =^\eval_{\l}\fdnu$, and $\Phi$ a distribution formula such that $\eval,\fdmu\not\models\Phi$ but $\eval, \fdnu\models\Phi$. We construct another distribution formula $\Phi'$ satisfying $\eval,\fdmu\models\Phi'$ but $\eval, \fdnu\not\models\Phi'$ by induction on the structure of $\Phi$.
\begin{itemize}
\item[(i)] $\Phi=Q_{\gtrsim \a}(\phi)$.  Let $$S=\{\snu\in SN : \eval, \snu\models \phi\}\hspace{2em}\mbox{  and  }\hspace{2em} \overline{S}=SN-S.$$ Then by definition, $\fdnu(S)\gtrsim \a$ but $\fdmu(S)\not\gtrsim \a$. 
Let $\b=\fdmu(\overline{S})$ and  $\Phi'=Q_{\gtrsim \b}(\neg\phi)$. Then we have trivially $\eval, \fdmu\models \Phi'$. Now it suffices to show
$\eval, \fdnu\not\models \Phi'$. Otherwise, we have $\fdnu(\overline{S})\gtrsim\b$, and then
$$\id_\h \eqsim \fdnu(S) + \fdnu(\overline{S}) \gtrsim \a + \b.$$
On the other hand, we have
$$\id_\h \eqsim \fdmu(S) + \fdmu(\overline{S}) = \fdmu(S) + \b.$$
Comparing the two formulae above, we conclude that $\fdmu(S)\gtrsim \a$, a contradiction. 
\item[(ii)] $\Phi=\bigwedge_{i\in I} \Phi_i$. Then by definition, $\eval, \fdnu\models\Phi_i$ for each $i\in I$ but $\eval, \fdmu\not\models\Phi_{i_0}$ for some $i_0\in I$.  
By induction we have $\Phi_{i_0}'$ such that $\eval, \fdmu\models \Phi_{i_0}'$ but $\eval, \fdnu\not\models \Phi_{i_0}'$. 
For any $i\not =i_0$, let $\Phi_i'=\Phi_i$ if $\eval, \fdmu\models\Phi_i$, and otherwise it is determined by applying induction on $\Phi_i$.
Let $\Phi'=\bigwedge_{i\in I} \Phi'_i$. 
Then $\eval, \fdmu\models \Phi'$ but $\eval, \fdnu\not\models \Phi'$.
\hfill $\Box$ 
\end{itemize}

With this lemma, we can show that the logic $\l$ exactly characterises the behaviours of quantum snapshots up to symbolic bisimilarity.

\begin{theorem}
Let $\snt$ and $\snu$ be two snapshots and $b\in BExp$. Then $\snt\bis^b\snu$ if and only if for any evaluation $\eval$, $\eval(b)=\true$ implies $\snt=^\eval_\l\snu$. 
\end{theorem}
{\it Proof.} We first prove the necessity part. For any $\phi, \Phi\in \l$, it suffices to prove the following two properties:
\begin{eqnarray*}
&& \forall\ \snt,\snu,\eval, \mbox{ if  $\snt\bis^b\snu$ and $\eval(b)=\true$ then }\eval,\snt\models \phi \Leftrightarrow \eval,\snu\models\phi,\\
&& \forall\ \fdmu, \fdnu, \eval, \mbox{ if  $\fdmu\bis^b\fdnu$ and $\eval(b)=\true$ then }\eval,\fdmu\models \Phi \Leftrightarrow \eval,\fdnu\models\Phi.
\end{eqnarray*}
We proceed by mutual induction on the structures of $\phi$ and $\Phi$. Take arbitrarily $\snt\bis^b\snu$, $\fdmu\bis^b\fdnu$, and $\eval(b)=\true$. Let 
$\snt = \osnt$, $\snu = \osnu$, $\eval,\snt\models \phi$, and $\eval,\fdmu\models \Phi$. There are seven cases to consider:
\begin{itemize}
\item $\phi=\g_{\widetilde{q}}$. Then $qv(\ott)\cap \widetilde{q} = \emptyset$ and $\e\eqsim_{\widetilde{q}} \g$.
Since $\snt\bis^b\snu$ and $b$ is satisfiable, we have $qv(\ott)= qv(\otu)$ and $\e\eqsim_{\overline{qv(\ott)}} \f$. Thus $qv(\otu)\cap \widetilde{q} = \emptyset$, and 
$\f\eqsim_{\widetilde{q}} \g$ from the fact that $\widetilde{q}\subseteq \overline{qv(\ott)}$. Then $\eval, \snu\models \g_{\widetilde{q}}$ follows.

\item $\phi=  \neg \phi'$. Then $\eval, \snt \not\models \phi'$. By induction we have  $\eval, \snu \not \models \phi'$, and  $\eval, \snu \models \phi$.

\item $\phi=  \bigwedge_{i\in I} \phi_i$. Then $\eval, \snt \models \phi_{i}$ for each $i\in I$. By induction we have  $\eval, \snu \models \phi_{i}$, and  $\eval, \snu \models \phi$. 

\item $\phi=  \g.\phi'$. Then $\g\in CP_t(\h_{\overline{qv(\ott)}})$ and $\eval, \g(\snt)  \models \phi'$. Since $\snt\bis^b \snu$, we have $\g(\snt)\bis^b \g(\snu)$ by Proposition~\ref{prop:biscg}, and $qv(\ott)=qv(\otu)$. By induction we have $\eval, \g(\snu) \models \phi'$, and  $\eval, \snu \models \phi$.

\item $\phi= \<\sact\>\Phi'$. Then $\snt \rto{b_1, \sact'}\fdmu'$ for some $b_1$, $\sact'$, and $\fdmu'$ such that $\eval(b_1)=\true$, $\sact=_\eval \sact'$, and $\eval, \fdmu'\models \Phi'$. 
Since $\snt\bis^b \snu$,
there exists a collection of booleans $B$ such that $b\wedge b_1\ra \bigvee B$ and $\forall\ b'\in B$, $\exists b(b'), \sact(b')$ with $b'\ra b(b')$,
$\sact'=_{b'} \sact(b')$, $\su\rto{b(b'), {\sact(b')}} \fdnu'$, and $\fdmu' \bis^{b'} \fdnu'$. Note that $\eval(b\wedge b_1)=\true$. We can find a $b'\in B$ such that $\eval(b')=\true$. Thus $\eval(b(b'))=\true$, and $\sact=_{\eval} \sact(b')$.
Furthermore, by induction we have $\eval, \fdnu'\models \Phi'$ from $\fdmu' \bis^{b'} \fdnu'$ and $\eval, \fdmu'\models \Phi'$. So $\eval, \snu \models  \<\sact\>\Phi'$.

\item $\Phi= Q_{\gtrsim \a}(\phi')$. Let $S=\{\snt\in SN : \eval, \snt\models \phi'\}$. Then by definition, $\fdmu(S)\gtrsim\a$.
Furthermore, by induction we can see that $S$ is the disjoint union of some equivalence classes $S_1, \cdots, S_k$ of $\bis^{b}$. Thus 
$$\fdnu(S)=\fdnu(S_1)+\cdots + \fdnu(S_k)\eqsim\fdmu(S_1)+\cdots + \fdmu(S_k)=\fdmu(S)\gtrsim \a$$
where the $\eqsim$ equality is derived from the assumption that $\fdmu \bis^{b} \fdnu$.

\item $\Phi=  \bigwedge_{i\in I} \Phi_i$. Then $\eval, \fdmu \models \Phi_{i}$ for each $i\in I$. By induction we have  $\eval, \fdnu \models \Phi_{i}$, and  $\eval, \fdnu \models \Phi$. 
\end{itemize}
By symmetry, we also have $\eval,\snu\models \phi$ implies $\eval,\snt\models \phi$ and $\eval,\fdnu\models \Phi$ implies $\eval,\fdmu\models \Phi$.
That completes the proof of the necessity part.

We now turn to the sufficiency part. By Lemma~\ref{lem:sr}, we need only to prove that $\snt=^\eval_\l\snu$ implies 
$(\st\eval)(\rho)\stbis (\su\eval)(\rho)$ for all $\rho\in \dh$.
Let
$$\r = \{((\st\eval)(\rho), (\su\eval)(\rho)) : \rho\in \dh, \eval\in EV, \mbox{ and }\snt=^\eval_\l\snu\}$$
It suffices to show that $\r$ is an open bisimulation. Suppose
$(\st\eval)(\rho)\r (\su\eval)(\rho)$. Then $\snt=^\eval_{\l}\snu$, and
$$qv(\ott\eval)=qv(\ott)=qv(\otu)=qv(\otu\eval).$$
We further claim that $\tr_{qv(\ott)}\e(\rho) = \tr_{qv(\ott)}\f(\rho)$. Otherwise there exists $\widetilde{q}\subseteq \overline{qv(\ott)}$ such that $\e \not\eqsim_{\widetilde{q}}\f$. Then $\eval, \snt \models \e_{\widetilde{q}}$ while $\eval, \snu \not\models \e_{\widetilde{q}}$, a contradiction. 
 
Now let $(\st\eval)(\rho)\srto{\alpha}\dmu$. By Lemma~\ref{lem:so2s}
we have $\snt\rto{b_1, \sact}\fdmu_\dmu$ such that $\eval(b_1)=\true$, $\dmu=(\fdmu_\dmu\eval')(\rho)$, and
\begin{enumerate}
\item if $\alpha=c?v$ then $\sact=c?x$ for some $x\not\in fv(\ott)$, and $\eval'=\eval\{v/x\}$,
\item
otherwise, $\sact=_\eval\alpha$ and $\eval'=\eval$.
\end{enumerate}
Let $$\k = \{\dnu\in \dist(Con) : (\su\eval)(\rho)\srto{\alpha}\dnu \mbox{ and } \dmu\not\!\r\dnu\}.$$
For any $\dnu\in \k$, by Lemma~\ref{lem:so2s}
we have $\snu\rto{b(\fdnu_\dnu), \sact(\fdnu_\dnu)}\fdnu_\dnu$ such that $\eval(b(\fdnu_\dnu))=\true$, $\dnu=(\fdnu_\dnu\eval'')(\rho)$, and
\begin{enumerate}
\item if $\alpha=c?v$ then $\sact(\fdnu_\dnu)=c?x$ for some $x\not\in fv(\otu)$, and $\eval''=\eval\{v/x\}$,
\item
otherwise, $\sact(\fdnu_\dnu)=_\eval\alpha$ and $\eval''=\eval$.
\end{enumerate}
Here again, to ease the notations we only consider the case where for each $\fdnu$, there is at most one pair, denoted $(b(\fdnu), \sact(\fdnu))$, such that $\snu \rto{b(\fdnu), \sact(\fdnu)}\fdnu$. Furthermore, by $\alpha$-conversion, we can always take $\sact(\fdnu_\nu)=_{\eval}\sact$ and $\eval''=\eval'$.
For any $\dnu\in \k$, we claim $\fdmu_\dmu\not=^\eval_\l \fdnu_\dnu$. Otherwise, since $\dmu=(\fdmu_\dmu\eval')(\rho)$ and $\dnu=(\fdnu_\dnu\eval')(\rho)$, we have $\dmu\r\dnu$, a contradiction.
Thus, from Lemma~\ref{lem:witness} (2), there exists $\Phi_\dnu\in \l$ such that $\eval, \fdmu_\dmu\models \Phi_\dnu$ but $\eval, \fdnu_\dnu\not\models \Phi_\dnu$.
Let $$\Phi_{\dmu} = \bigwedge\{ \Phi_{\dnu} : \dnu\in \k\}\mbox{  and  } \phi=\<\sact\>\Phi_{\dmu}.$$
Then $\eval, \fdmu_{\dmu}\models \Phi_{\dmu}$ and $\eval, \snt\models \phi$.
Since $\snt=^\eval_{\l}\snu$, we have $\eval, \snu\models \phi$ too. That is, there exists $\fdomega$ such that $\eval(b(\fdomega))=\true$, $\sact=_\eval \sact(\fdomega)$, and $\eval,\fdomega\models \Phi_{\dmu}$. Now by Lemma~\ref{lem:ss2o}, we have $ (\su\eval)(\rho)\srto{\alpha'}\domega=(\fdomega\eval''')(\rho)$ such that 
\begin{enumerate}
\item
if $\sact(\fdomega)=c?x$ then $\alpha'=c?v$ for some $v\in \qc{Real}$, and $\eval'''=\eval\{v/x\}$,
\item
otherwise, $\alpha'=_\eval\sact(\fdomega)$ and $\eval'''=\eval$.
\end{enumerate}
By transition rule $C$-$Inp_c$, we can alway choose $\alpha'=\alpha$, and $\eval'''=\eval'$.
We claim that $\domega\not\in \k$. Otherwise, if $\domega\in \k$ then 
$\eval, \fdnu_\domega\not\models \Phi_\domega$, and $\eval, \fdnu_\domega\not\models \Phi_\dmu$ as well. This is a contradiction
since by assumption, $\fdnu_\domega =\fdomega$. 
So $\domega\not\in \k$, and $\dmu\r\domega$ as required.

Finally, we prove that $\r$ is closed under
super-operator application. To this end, we only need to show that $=^\eval_\l$ is
; that is, for any
$\g\in CP_t(\h_{\overline{qv(\ott)}})$, $\snt =^\eval_\l
\snu$ implies
$\g(\snt)=^\eval_\l \g(\snu)$.
Suppose $\snt=^\eval_\l\snu$ and let $\phi$ be a formula such that
$\eval, \g(\snt)\models\phi$. Then $\eval,\snt\models\g.\phi$. It follows from
$\snt =^\eval_\l \snu$ that $qv(\ott)=qv(\otu)$ and $\eval,\snu\models
\g.\phi$. Therefore, $\eval, \g(\snu)\models\phi$. By symmetry if
$\phi$ is satisfied by $\eval,\g(\snu)$ then it is also satisfied by
$\eval,\g(\snt)$. In other words, we have $\g(\snt)=^\eval_\l \g(\snu)$.
Then $\r$
is an open bisimulation by Proposition 5 of \cite{DF11}. \hfill $\Box$

 For any $\ott, \otu\in \t$ and $b\in BExp$,  we write $\ott =^b_{\l}\otu$ if  for any evaluation $\eval$, $\eval(b)=\true$ implies
$\sl \ott, \id_\h\sr =^\eval_{\l}\sl \otu, \id_\h\sr$. Then we have the following theorem:
 
\begin{theorem}
For any $\ott, \otu\in \t$, $\ott\bis^b \otu$ if and only if $\ott =^b_{\l}\otu$.
\end{theorem}

\section{Conclusion and further work}

The main contribution of this paper is a notion of symbolic bisimulation for qCCS, a quantum extension of classical value-passing CCS. 
By giving the operational semantics of qCCS directly by means of the super-operators a process can perform, we are able to
assign to each (non-recursively defined) quantum process a $finite$ super-operator weighted labelled transition system, comparing to the $infinite$ probabilistic labelled transition system in previous literature.
We prove that the symbolic bisimulation in this paper coincides with the open bisimulation in~\cite{DF11}, thus providing a practical way to decide the latter. We also 
design an algorithm to check symbolic ground bisimulation, which is applicable to reasoning about the correctness of existing quantum communication protocols. 
A modal logic characterisation for the symbolic bisimulation is also developed.

A natural extension of the current paper is to study symbolic weak bisimulation where the invisible actions, caused by internal  (classical and quantum)
communication as well as quantum operations, are abstracted away. To achieve this, we may need to define symbolic weak transitions similar to those proposed
in~\cite{FDY11} and \cite{DF11}. Note that one of the distinct features of weak transitions for probabilistic processes is the so-called left decomposibility; that is,
 if $\dmu\Rto{}\dnu$ and $\dmu=\sum_{i\in I} p_i \dmu_i$ is a probabilistic decomposition of $\dmu$, then $\nu$ can be decomposed into $\sum_{i\in I}p_i \dnu_i$ accordingly such that $\dmu_i\Rto{}\dnu_i$ for each $i\in I$. This property is useful in proving the transitivity of bisimilarity. However, it is not satisfied by symbolic transitions defined in this paper, since, in general, a super-operator does not have an inverse. Therefore, we will have to explore other ways of defining weak symbolic transitions, which is one of the research directions we are now pursing.
 
We have presented in this paper, for the first time in literature to the best of our knowledge, the notion of super-operator weighted labelled transition systems, which serves the semantic model for qCCS and plays an important role in describing and reasoning about quantum processes. For the next step, we are going to explore the possibility of model checking quantum communication protocols based on this model. As is well known, one of the main challenges for quantum model checking is 
that the set of all quantum states, traditionally regarded as the underlying state space of the models to be checked, is a continuum, so that the techniques of classical model checking, which normally works only for finite state space, cannot be applied directly. Gay et al.~\cite{GNP06, GNP08, Pa08} provided a solution for this problem by restricting the state space to a set of finitely describable states called stabiliser states, and restricting the quantum operations applied on them to the class of Clifford group. By doing this, they were able to obtain an efficient model checker for quantum protocols, employing purely classical algorithms. 
The limit of their approach is obvious: it can only check the (partial) behaviours of a protocol on stabiliser states, and
does not work for general protocols. 

Our approach of treating both classical data and quantum operations in a symbolic way provides  an efficient and compact way to describe behaviours of a quantum protocol without resorting to the underlying quantum states. In this model, all existing quantum protocols have finite state spaces, and consequently, classical model checking techniques will be easily adapted to verifying quantum protocols.
 
 \section*{Acknowledgement}
This work was supported by Australian ARC grants DP110103473 and FT100100218.
 
\bibliographystyle{plain}
\bibliography{allinone}

\begin{thebibliography}{10}

\bibitem{BB84}
C.~H. Bennett and G.~Brassard.
\newblock Quantum cryptography: Public-key distribution and coin tossing.
\newblock In {\em Proceedings of the IEEE International Conference on Computer,
  Systems and Signal Processing}, pages 175--179, 1984.

\bibitem{BB93}
C.~H. Bennett, G.~Brassard, C.~Crepeau, R.~Jozsa, A.~Peres, and W.~Wootters.
\newblock Teleporting an unknown quantum state via dual classical and epr
  channels.
\newblock {\em Physical Review Letters}, 70:1895--1899, 1993.

\bibitem{BW92}
C.~H. Bennett and S.~J. Wiesner.
\newblock Communication via one- and two-particle operators on
  einstein-podolsky-rosen states.
\newblock {\em Physical Review Letters}, 69(20):2881--2884, 1992.

\bibitem{BC92}
J.R. Burch, E.M. Clarke, K.L. McMillan, D.L. Dill, and L.J. Hwang.
\newblock Symbolic model checking: $10^{20}$ states and beyond.
\newblock {\em Information and Computation}, 98(2):142--170, 1992.

\bibitem{Da11}
T.~A.~S. Davidson.
\newblock {\em Formal Verification Techniques using Quantum Process Calculus}.
\newblock PhD thesis, 2011.

\bibitem{DF11}
Yuxin Deng and Yuan Feng.
\newblock Open bisimulation for quantum processes.
\newblock Manuscript. Available at http://arxiv.org/abs/1201.0416.

\bibitem{FDJY07}
Y~Feng, R~Duan, Z~Ji, and M~Ying.
\newblock Probabilistic bisimulations for quantum processes.
\newblock {\em Information and Computation}, 205(11):1608--1639, 2007.

\bibitem{FDY11}
Y~Feng, R~Duan, and M~Ying.
\newblock Bisimulations for quantum processes.
\newblock In Mooly Sagiv, editor, {\em Proceedings of the 38th ACM Symposium on
  Principles of Programming Languages (POPL'11)}, pages 523--534, 2011.

\bibitem{GNP06}
S~Gay, R~Nagarajan, and N~Papanikolaou.
\newblock Probabilistic model-checking of quantum protocols.
\newblock In {\em Proceedings of the 2nd International Workshop on Developments
  in Computational Models}, 2006.

\bibitem{GNP08}
S~Gay, R~Nagarajan, and N~Papanikolaou.
\newblock Qmc: A model checker for quantum systems.
\newblock In {\em CAV ’08}, pages 543--547. Springer, 2008.

\bibitem{GN05}
S.~J. Gay and R.~Nagarajan.
\newblock Communicating quantum processes.
\newblock In J.~Palsberg and M.~Abadi, editors, {\em Proceedings of the 32nd
  ACM SIGPLAN-SIGACT Symposium on Principles of Programming Languages (POPL)},
  pages 145--157, 2005.

\bibitem{HL95}
M.~Hennessy and H.~Lin.
\newblock Symbolic bisimulations.
\newblock {\em Theoretical Computer Science}, 138(2):353--389, 1995.

\bibitem{JL04}
P.~Jorrand and M.~Lalire.
\newblock Toward a quantum process algebra.
\newblock In P.~Selinger, editor, {\em Proceedings of the 2nd International
  Workshop on Quantum Programming Languages, 2004}, page 111, 2004.

\bibitem{Kr83}
K.~Kraus.
\newblock {\em States, Effects and Operations: Fundamental Notions of Quantum
  Theory}.
\newblock Springer, Berlin, 1983.

\bibitem{La06}
Marie Lalire.
\newblock Relations among quantum processes: Bisimilarity and congruence.
\newblock {\em Mathematical Structures in Computer Science}, 16(3):407--428,
  2006.

\bibitem{NC00}
M.~Nielsen and I.~Chuang.
\newblock {\em Quantum computation and quantum information}.
\newblock Cambridge university press, 2000.

\bibitem{Pa08}
N.~K. Papanikolaou.
\newblock {\em Model Checking Quantum Protocols}.
\newblock PhD thesis, 2008.

\bibitem{San96}
D.~Sangiorgi.
\newblock A theory of bisimulation for the œÄ-calculus.
\newblock {\em Acta Informatica}, 33(1):69--97, 1996.

\bibitem{vN55}
J.~von Neumann.
\newblock {\em Mathematical Foundations of Quantum Mechanics}.
\newblock Princeton University Press, Princeton, NJ, 1955.

\bibitem{YFDJ09}
M~Ying, Y~Feng, R~Duan, and Z~Ji.
\newblock An algebra of quantum processes.
\newblock {\em ACM Transactions on Computational Logic (TOCL)}, 10(3):1--36,
  2009.

\end{thebibliography}

\end{document}